\documentclass[
  aps,
  prx,  
  reprint,
  groupedaddress,
]{revtex4-2}

\usepackage{microtype}
\usepackage[svgnames]{xcolor}
\usepackage[hidelinks]{hyperref}
\usepackage{mathtools}
\usepackage{bm}
\usepackage[normalem]{ulem}

\usepackage{csquotes}

\usepackage{newtxtext}
\usepackage{newtxmath}

\newcommand*{\HeII}{He~II}
\newcommand*{\Lint}{L_0}
\newcommand*{\Lbox}{L_{\text{box}}}
\newcommand*{\Tsim}{T_{\text{sim}}}
\newcommand*{\dtSnap}{\Delta t_{\text{snap}}}
\newcommand*{\Tint}{T_0}
\newcommand*{\azero}{\ensuremath{a_0}}
\newcommand*{\Ccal}{\mathcal{C}}
\newcommand*{\Scal}{\mathcal{S}}  
\newcommand*{\Ecal}{\mathcal{E}}
\newcommand*{\Fcal}{\mathcal{F}}
\newcommand*{\Lcal}{\mathcal{L}}  
\newcommand*{\Cint}{\oint_{\Ccal}}  
\newcommand*{\dd}{\mathrm{d}}
\newcommand*{\svec}{\bm{s}}
\newcommand*{\xvec}{\bm{x}}
\newcommand*{\rvec}{\bm{r}}
\newcommand*{\kvec}{\bm{k}}
\newcommand*{\vvec}{\bm{v}}
\newcommand*{\vortvec}{\bm{\omega}}
\newcommand*{\vortsvec}{\vortvec_{\text{s}}}
\newcommand*{\psivec}{\bm{\psi}}
\newcommand*{\psisvec}{\psivec_{\text{s}}}
\newcommand*{\vs}{\vvec_{\text{s}}}  
\newcommand*{\vL}{\vvec_{\text{L}}}  
\newcommand*{\vf}{\vvec_{\text{f}}}  
\newcommand*{\kf}{k_{\text{f}}}  
\newcommand*{\kell}{k_{\ell}}  
\newcommand*{\vshat}{\widehat{\vvec}_{\text{s}}}  
\newcommand*{\vortsshat}{\widehat{\vortvec}_{\text{s}}}  
\newcommand*{\grad}{\bm{\nabla}}
\newcommand*{\curl}{\grad \times}
\newcommand*{\laplacian}{\nabla^2}
\newcommand*{\eps}{\varepsilon}
\newcommand*{\epsInj}{\eps_{\text{inj}}}
\newcommand*{\epsSigma}{\eps_{\sigma}}

\newcommand*{\deltaRes}{\delta_{\text{res}}}
\newcommand*{\ReyKappa}{\ensuremath{\textit{Re}_\kappa}}
\newcommand*{\vrms}{\ensuremath{v_{\text{rms}}}}
\newcommand*{\near}{^{\text{(n)}}}
\newcommand*{\far}{^{\text{(f)}}}

\newcommand*{\rcut}{r_{\text{cut}}}
\newcommand*{\kmax}{k_{\text{max}}}
\newcommand*{\Vmean}[1]{{\langle #1 \rangle}_{V}}  
\newcommand*{\Lmean}[1]{{\langle #1 \rangle}}  
\newcommand*{\Csigma}{C_\sigma}

\DeclareMathOperator{\erf}{erf}
\DeclareMathOperator{\erfc}{erfc}

\renewcommand*{\paragraph}[1]{\textit{#1---}}

\begin{document}


\title{Nonlocal Energy Transfer Mechanism in Three-dimensional Quantum Turbulence}


\author{Elliot Bes}
\author{Guillaume Balarac}
\author{Juan Ignacio Polanco}
\email[]{juan-ignacio.polanco@cnrs.fr}
\affiliation{Univ.\ Grenoble Alpes, CNRS, Grenoble INP, LEGI, 38000 Grenoble, France}


\date{\today}

\begin{abstract}
  We investigate the kinetic energy cascade in zero-temperature quantum turbulence.
  Using simple theoretical arguments and unprecedented numerical simulations,
  we unveil an universal mechanism transferring energy directly from large to very small scales,
  thus bypassing the Kolmogorov-like local energy cascade and resulting in
  steepening of the energy spectra at scales larger than the typical distance between quantum vortices.
  This mechanism rests both on the vast separation of
  scales typical of superfluid helium-4 flows and on the alignment
  between quantum vortices and large-scale velocity gradients,
  in direct analogy with vortex stretching in classical turbulent flows.
\end{abstract}


\maketitle



\paragraph{Introduction}%
Near absolute zero, helium-4 exists in a superfluid state known as \HeII{}
which has remarkable hydrodynamical properties.
In addition to being inviscid, all vorticity of zero-temperature \HeII{} is
concentrated in atomically thin vortices with quantized circulation
$\kappa \approx 0.997 \, 10^{-7} \, \text{m}^2/\text{s}$ known as quantum
vortices~\cite{Donnelly1991}.
Despite these major differences with classical fluids, quantum
turbulence (QT) in \HeII{}~\cite{Barenghi2023} is widely thought to present
significant similarities with classical turbulence, especially at large scales where the
collective behavior of quantum vortices becomes dominant~\cite{Maurer1998,Baggaley2012d,Muller2021,Bret2025,Polanco2025a,Galantucci2025}.
In typical \HeII{} experiments~\cite{Maurer1998,Walmsley2007,Paoletti2008a,Svancara2017,Tang2020,Salort2021,Bret2025}, QT
is characterized by a very wide range of scales, ranging from the microscopic vortex
core radius $\azero \sim 10^{-10}\,\text{m}$ to the integral scale
$\Lint \sim 10^{-2}\text{--}10^{-1}\,\text{m}$ associated to the largest turbulent structures.
Between the two, the typical distance between quantum vortices, $\ell \sim 10^{-5}\text{--}10^{-4}\,\text{m}$,
sets the transition between
the \emph{quantum} scales, where the discrete nature of quantum vortices is apparent,
and the \emph{inertial} scales, where the collective motion and the partial
polarization of quantum
vortices~\cite{Lvov2007,Roche2008,Baggaley2011a,Polanco2021} can lead to
quasi-classical behavior.

Up to now, the inertial and quantum ranges have been mostly regarded as two
disconnected regions of scale space with their own very different dynamics~\cite{Vinen2002,Lvov2007}.
In this Letter we show that this is not truly the case, since, due to the vast
separation between $\ell$ and $\azero$ in \HeII{} turbulence, the quantum range
acts as an energy sink which directly receives kinetic energy from the inertial
scales, thus bypassing the locality of the Richardson cascade on which rests
Kolmogorov's classical turbulence phenomenology~\cite{Frisch1995}.
A major consequence is that, even at scales significantly larger than $\ell$, the kinetic energy spectrum significantly departs from
the expected quasi-classical prediction $E(k) \sim \varepsilon^{2/3} k^{-5/3}$, where $k$ is the wavenumber
and $\varepsilon$ the (constant) energy flux through scale $\sim 1/k$.
Similar \emph{spectral shortcut} mechanisms have been identified in classical three-dimensional (3D) turbulent flows
interacting with plant canopies~\cite{Finnigan2000} and rigid slender fibers~\cite{Olivieri2020,Cannon2024},
as well as in geophysical quasi-two-dimensional settings where turbulent
fluctuations can feed the mean flow~\cite{Srinivasan2012,Marston2023}.

Using simple theoretical arguments, we show that in QT such nonlocal energy transfers rest on the preferential
alignment of quantum vortices with large-scale straining motions induced by
the vortex tangle itself.
Such alignment tends to increase the quantum vortex line density $\Lcal \equiv \ell^{-2}$
and thus the energy content $E_{\text{small}} \sim \kappa^2 \Lcal \ln ( \ell / \azero )$ at quantum scales,
at the expense of large-scale energy.
Note that, unlike energy, vortex length is not a conserved quantity and therefore it is allowed to increase.
This vortex line production mechanism can be directly related to the ubiquitous vortex stretching
phenomenon in classical turbulent
flows~\cite{Tennekes1972,Ashurst1987a,Buaria2020c}, which arises from similar alignments.
The identified mechanism appears to be a generic property of QT as well, as it can be deduced from
past numerical~\cite{Nore1997,Araki2002,Villois2016a,Muller2020} and
experimental~\cite{Stalp1998,Barenghi2006,Walmsley2007,Walmsley2014} observations of growth of vortex
length at early times in decaying settings.
While these observations have been broadly associated to a transfer of energy to small
scales by the Richardson cascade~\cite{Walmsley2014}, the precise mechanism
explaining them remained unidentified until now.
At later times, as a disordered vortex tangle develops, such vortex growth is likely to be
arrested by the loss of vortex length associated to vortex
reconnections~\cite{Leadbeater2001,Villois2020}.
In particular, in steady-state regimes, the identified line production mechanism
can be seen as a necessary condition for preserving the total vortex line density over
time.

In the following, we describe quantum vortex dynamics at scales much larger than $\azero$
using the vortex filament model (VFM)~\cite{Schwarz1985,Hanninen2014}.
In the VFM, quantum vortices are described as closed or infinite lines in 3D space,
parametrized by the vortex positions $\svec(\xi)$ where $\xi$ is the arc length.
A tangle of quantum vortices induces a velocity field through space given by the Biot--Savart law,
\begin{equation}
  \vs(\xvec, t) = \frac{\kappa}{4\pi} \oint_{\Ccal(t)} \! \frac{(\svec - \xvec) \times \dd \svec}{|\svec - \xvec|^3},
  \label{eq:BiotSavart}
\end{equation}
where $\Ccal(t)$ denotes the instantaneous vortex geometry.
When evaluated on a vortex location $\svec_0 \in \Ccal(t)$, integration must be omitted over a small distance
$|\svec - \svec_0| < \azero \, e^{\Delta}/2$ (we set $\Delta = 1/2$~\cite{Hanninen2014})
to avoid the singularity in Eq.~\eqref{eq:BiotSavart} and to account
for the finite vortex core size $\azero$~\cite{Saffman1993}.
A well known result of this desingularization procedure
is the localized induction approximation (LIA)~\cite{Arms1965},
\begin{equation}
  \vs^{\epsilon}(\svec) = \frac{\kappa}{4\pi} \svec' \times \svec'' \, \ln\left( \frac{2\epsilon}{e^\Delta \azero} \right),
  \quad a_0 \ll \epsilon \ll \frac{1}{|\svec''|},
  \label{eq:LIA}
\end{equation}
which expresses the velocity induced by a small local segment $\xi \in [\xi_0 - \epsilon, \xi_0 + \epsilon]$
on a vortex point $\svec(\xi_0)$, and primes denote derivatives with respect to $\xi$.
In particular, $\svec'$ and $\svec''$ are respectively the local unit tangent and curvature vectors.

\paragraph{Spectral energy fluxes}%
In the following we consider a statistically isotropic and spatially
homogeneous turbulent quantum vortex tangle in a cubic periodic domain of
volume $V = \Lbox^3$.
The kinetic energy per unit mass associated to the velocity field~\eqref{eq:BiotSavart} is
$E = \Vmean{|\vs|^2} / 2 = \frac{\kappa^2}{8\pi V} \Cint \Cint \frac{\dd \svec_1 \cdot \dd \svec_2}{|\svec_1 - \svec_2|}$~\cite{Saffman1993},
where $\Vmean{\cdot}$ denotes a volume average.
The Biot--Savart law~\eqref{eq:BiotSavart} is then equivalent to~\cite{Sonin1987}
\begin{equation}
  \frac{\delta E}{\delta \svec} = \frac{\kappa}{V} \svec' \times \vs(\svec),
  \label{eq:magnus_force}
\end{equation}
which is interpreted as a Magnus force (per unit length and mass) acting on each vortex location $\svec$.
From the definition of the functional derivative $\frac{\delta E}{\delta \svec}$, the
energy variation rate due to an arbitrary vortex velocity $\vL(\svec)$ is
$\left. \frac{\dd E}{\dd t} \right|_{\vL} = \Cint \frac{\delta E}{\delta \svec} \cdot \vL \, \dd \xi$.

By decomposing $\vs(\xvec)$ into Fourier modes,
one can readily generalize Eq.~\eqref{eq:magnus_force} to
$\frac{\delta \Ecal_k}{\delta \svec} = \frac{\kappa}{V} \svec' \times \vs^{<k}(\svec)$,
where $\vs^{<k}(\xvec) = \sum_{|\kvec| \le k} \vshat(\kvec) \, e^{i \kvec \cdot \xvec}$
is a low-pass filtered velocity field
and $\Ecal_k = \Vmean{|\vs^{<k}|^2} / 2$ is the cumulative
energy up to wavenumber $k$~\cite{Frisch1995}.
Ultimately, this allows us to define the energy flux through wavenumber $k$ as
the energy removed from the large scales by the total induced velocity $\vs$~\eqref{eq:BiotSavart},
\begin{equation}
  \Pi_k
    \equiv - \left. \frac{\dd \Ecal_k}{\dd t} \right|_{\vs} \!\!
    = - \Cint \frac{\delta \Ecal_k}{\delta \svec} \cdot \vs \, \dd \xi
    = \frac{\kappa}{V} \Cint \left( \svec' \times \vs \right) \cdot \vs^{<k} \, \dd \xi.
  \label{eq:energy_flux}
\end{equation}
Since the superfluid vorticity can be formally expressed as $\vortsvec(\xvec) = \kappa \Cint \delta(\xvec - \svec) \, \dd \svec$,
this is in fact equal to the usual definition
$\Pi_k = \Vmean{\vs^{<k} \cdot (\vortsvec \times \vs)} = \Vmean{\vs^{<k} \cdot (\vs \cdot \grad \vs)}$
in incompressible Navier--Stokes turbulence~\cite{Frisch1995}.
Note that, as in classical fluids, the flux vanishes as $k \to \infty$.

\paragraph{Nonlocal energy transfer}%
In 3D classical isotropic turbulence, for $k$ in the inertial range, the energy flux $\Pi_k$
is constant and equal to the energy cascade rate $\eps$.
Moreover, the energy cascade is local in the sense that $\vs^{<k}$ mostly
transfers energy to wavenumbers $k' \lesssim 2k$~\cite{Kraichnan1966,Zhou1993a,Eyink2009,Cardesa2017}.
In the following we show that this is generally not the case in
zero-temperature QT, as energy is directly transferred from inertial scales
($k \ll k_\ell \equiv 2\pi / \ell$) to quantum scales $k' \gg k_\ell$.
Here the mean inter-vortex distance is $\ell = \Lcal^{-1/2}$ where
$\Lcal = \frac{1}{V} \Cint \dd\xi$ is the vortex line density.

We start by generalizing Eq.~\eqref{eq:energy_flux} to write the nonlocal energy
transfer from large ($<k$) to small ($>q$) scales as
$\Pi_k^{>q}
= \frac{\kappa}{V} \Cint (\svec' \times \vs^{>q}) \cdot \vs^{<k} \, \dd \xi$,
where $\vs^{>q} \equiv \vs - \vs^{<q}$ is a high-pass filtered (i.e.\ small-scale) velocity.
Since nonlocal vortex interactions are negligible at scales below $\ell$,
one can safely assume that $\vs^{>k_\ell}$ is well approximated by the
LIA~\eqref{eq:LIA} with $\epsilon \sim \ell$
(see the Supplemental Material~\cite{Supplemental} for a numerical verification).
This ultimately leads to
\begin{equation}
  \Pi_k^{>k_\ell}
  \approx \frac{\kappa^2}{4\pi} \ln\left( \frac{2\ell}{e^{\Delta} \azero} \right)
  \Scal_k, \quad
  \Scal_k \equiv -\frac{1}{V} \Cint \svec'' \cdot \vs^{<k} \, \dd\xi,
  \label{eq:energy_transfer_nonlocal}
\end{equation}
where we have used the geometrical relation $\svec' \times (\svec' \times \svec'') = -\svec''$.
Equation~\eqref{eq:energy_transfer_nonlocal} states that nonlocal energy
transfers from inertial to quantum scales are enabled by a possible anti-alignment
between the large-scale velocity $\vs^{<k}$ and the local vortex curvature $\svec''$,
and are amplified by the scale separation between $\ell$ and $\azero$.

Importantly, one can interpret $\Scal_k$ as the production of vortex line density $\Lcal$ by $\vs^{<k}$.
Indeed, using integration by parts,
\begin{equation}
  \Scal_k
  = \frac{1}{V} \Cint \sigma_k \, \dd \xi
    = \Lcal \, \Lmean{\sigma_k}
  \quad \text{with } \sigma_k \equiv \svec' \cdot \frac{\partial \vs^{<k}}{\partial \xi},
  \label{eq:vortex_stretching}
\end{equation}
where $\Lmean{\cdot}$ denotes an average over vortex lines.
Here
$\sigma_k$ represents
the local vortex line growth rate (per unit length) due to
$\vs^{<k}$,
and can also be expressed as $\sigma_k = \svec' \cdot \grad\vs^{<k} \cdot \svec'$.
Remarkably,
this last expression
is strongly reminiscent of the
enstrophy ($\Omega \equiv \Vmean{|\vortvec|^2} / 2$) production term from the Navier--Stokes equations, which writes
$P_\Omega = \Vmean{\vortvec \cdot \grad \vvec \cdot \vortvec} = \Vmean{\omega_i S_{ij} \omega_j}$
where $S_{ij} = (\partial_i v_j + \partial_j v_i) / 2$ is the strain-rate tensor~\cite{Tennekes1972}.
This term is generally
positive in 3D classical turbulence due to the vortex stretching mechanism~\cite{Taylor1938,Tennekes1972,Buaria2020c},
intimately linked to energy cascade dynamics.
In the case of Eq.~\eqref{eq:vortex_stretching}, $\Scal_k$ corresponds to vortex stretching by the \emph{nonlocal} (or large-scale)
strain rate tensor, whose most extensional eigenvector has
been shown to preferentially align with vorticity in classical turbulence~\cite{Hamlington2008a,Buaria2021a}.
Therefore, it is natural to expect $\Scal_k$ to be positive for $k \ll k_\ell$
if the large scales of QT have any similarities with classical fluid
turbulence.
Paradoxically, such similarities would result in
positive nonlocal energy transfers $\Pi_k^{>k_\ell}$
[Eq.~\eqref{eq:energy_transfer_nonlocal}], ultimately leading to nonclassical
scaling laws at scales larger than $\ell$.
This scenario is confirmed using numerical simulations below.

\paragraph{Numerical simulations}%
We perform VFM numerical simulations of statistically stationary zero-temperature QT at high vortex densities with an
unprecedented separation of scales between $\Lint$ and $\ell$.
To circumvent the $O(N^2)$ cost of directly computing far-ranged interactions
between $N$ vortex points, we adopt a novel fast Fourier transform (FFT)-based approach~\cite{Polanco2025b}
which enables the efficient evaluation of Eq.~\eqref{eq:BiotSavart} in $O(N \log N)$ time (see End Matter for details).
Moreover, to eliminate limitations associated to usual time-decaying configurations -- such
as statistical convergence issues and the influence of initial conditions -- we devise a novel energy injection
scheme to keep the system in a statistically stationary state.
Our approach mirrors standard practice in numerical classical turbulence studies.
Concretely, we introduce an external forcing velocity $\vf$ such that vortices move according to
\begin{equation}
  \frac{\dd \svec}{\dd t} = \vs(\svec) + \vf(\svec).
  \label{eq:ds_dt}
\end{equation}
As detailed in the End Matter [Eq.~\eqref{eq:forcing_velocity}], $\vf$ can be defined such that it injects kinetic
energy at the largest scales of the system (wavenumbers $k < \kf$) at rate $\epsInj$, while leaving
smaller scales unmodified (as verified in Fig.~\ref{fig:fluxes3200}, open circles).
We also ensure that $\vf \cdot \svec'' = 0$ so that $\vf$ does not directly modify
the total vortex length (and thus the energy at quantum scales).
Note that determining $\vf$ requires knowing the Fourier-filtered velocity $\vvec^{<\kf}$ at all times, which is
greatly facilitated by our FFT-based numerical method.

\begin{figure}
  \begin{center}
    \includegraphics[width=\columnwidth]{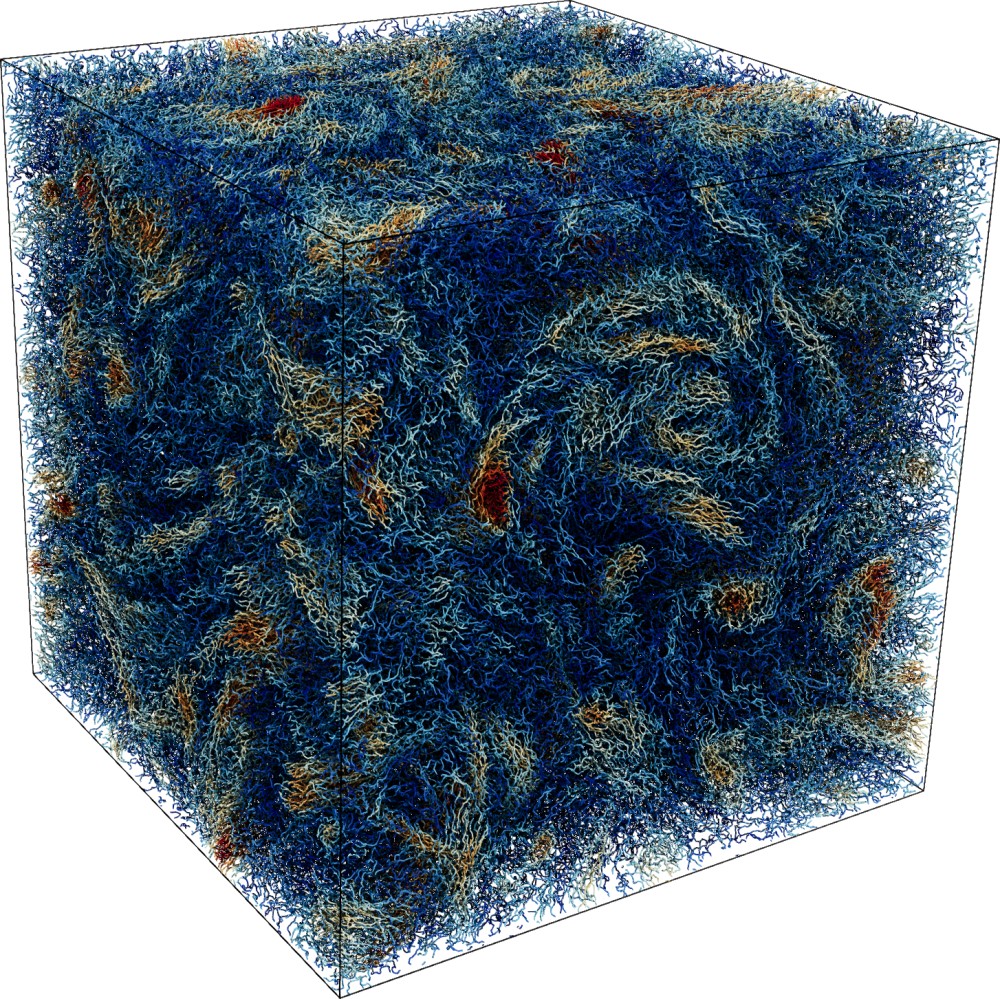}
  \end{center}
  \caption{\label{fig:visualisation}%
    Visualization of QT simulation in a statistically steady state (run $\epsInj = 3200$, $\azero = 10^{-7}$).
    Quantum vortices are colored by the Gaussian-filtered superfluid vorticity at scale $\ell$,
    highlighting highly polarized vortex bundles (yellow to red colors).
  }
\end{figure}

We set the nondimensional vortex circulation to $\kappa = 1$, the periodic domain size to
$\Lbox = 2\pi$ and the vortex core radius to $\azero = 10^{-7}$.
Since $\azero \approx 10^{-10}\,\text{m}$ in \HeII{}, this gives
$\Lbox \approx 2\pi\,\text{mm}$ in physical units, which is representative of a
typical measurement section in current experiments.
We also perform simulations with $\azero = 10^{-4}$ to illustrate the effect of reducing the $\ell/\azero$ ratio.
Forcing is applied up to wavenumber $\kf = 2.5$.
A total of 7 runs are performed with different values of $\azero$ and of the energy
injection rate $\epsInj$ as detailed in the End Matter (Table~\ref{tab:simulations}).
An instantaneous visualization of a typical run is shown in Fig.~\ref{fig:visualisation},
highlighting the formation of highly polarized vortex
bundles, a key feature of quasi-classical QT~\cite{Lvov2007,Roche2008,Baggaley2012,Baggaley2012c,Polanco2021}.

\begin{figure}[tb]
  \begin{center}
    \includegraphics{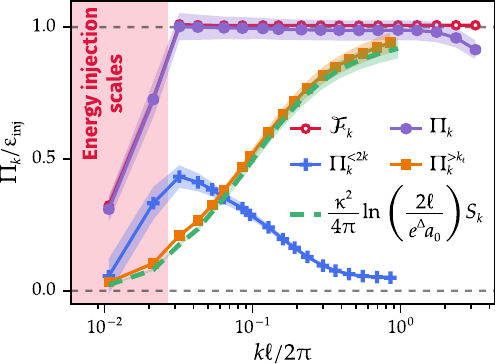}
  \end{center}
  \caption{\label{fig:fluxes3200}%
  Normalized spectral energy fluxes (run $\epsInj = 3200$, $\azero = 10^{-7}$).
  Accumulated energy injection rate $\Fcal_k$ (open circles, Eq.~\eqref{eq:forcing_injection_rate}),
  total energy flux $\Pi_k$ (filled circles, Eq.~\eqref{eq:energy_flux}),
  local energy flux $\Pi_k^{<2k}$ (crosses),
  nonlocal energy flux $\Pi_k^{>k_\ell}$ (squares),
  and estimated nonlocal flux based on vortex line production $\Scal_k$ (dashed line, Eq.~\eqref{eq:energy_transfer_nonlocal}).
  The energy injection scales ($k < \kf$) are highlighted in light red.
  Values are averaged across 31 flow snapshots within the statistically steady state.
  Light colored bands indicate 95\% confidence intervals (see End Matter for details).
  }
\end{figure}

\paragraph{Numerical results}%
We start by plotting in Fig.~\ref{fig:fluxes3200} the spectral energy fluxes obtained from one of our runs.
The total energy flux from large to small scales ($\Pi_k$, filled circles)
displays a plateau at scales larger than $\ell$ with a value close to the energy injection rate $\epsInj$.
At first sight, this is consistent with the energy cascade picture in classical turbulence~\cite{Alexakis2018}.
At smaller scales, the flux starts decreasing very slowly (logarithmically)
with $k$, and is expected to reach zero near the microscopic scale $k_a \sim 1 / \azero$
(not visible in the figure).
This drastically differs from the Navier--Stokes case where the energy flux
rapidly decays to zero when the viscous scales are reached.

The flux $\Pi_k$ provides no indication on the locality of the energy cascade.
With this aim, we consider the local flux $\Pi_k^{<2k}$
to neighboring scales $k' \in [k, 2k]$ (crosses in Fig.~\ref{fig:fluxes3200}), defined by replacing $\vs$ with $\vs^{<2k}$ in
Eq.~\eqref{eq:energy_flux}.
The local flux clearly decreases beyond the injection scales, thus breaking the
hypothesis of a constant energy flux associated to Kolmogorov's (K41) scaling laws.
This is also consistent with the behavior of the nonlocal flux to quantum scales
$\Pi_k^{>k_\ell}$ (squares), which is positive and increases throughout the inertial range.
  A more detailed analysis based on shell-to-shell energy transfers
  confirms this picture of a predominantly nonlocal energy transfer towards quantum scales~\cite{Supplemental}.
Besides, the nonlocal flux is in fairly good agreement with our estimation~\eqref{eq:energy_transfer_nonlocal}
based on the vortex length production term $\Scal_k$ (dashed line), providing strong support
to the proposed picture of a nonlocal energy transfer driven by the alignment
between quantum vortices and large-scale velocity gradients.
Our results confirm that a preferential alignment exists at all
scales (up to the largest simulated scale) leading to a continuous
depletion of the local cascade.
More profoundly, they also suggest that this is the main mechanism enabling a
transfer of energy from large to quantum scales, before
it being converted to acoustic energy (phonons) at microscopic scales via vortex reconnections~\cite{Nore1997a,Leadbeater2001,Kobayashi2005}
and the Kelvin wave cascade~\cite{Kivotides2001a,Vinen2003,Kozik2004,Lvov2010a,Krstulovic2012,Baggaley2014}.

\begin{figure}[tb]
  \begin{center}
    \includegraphics{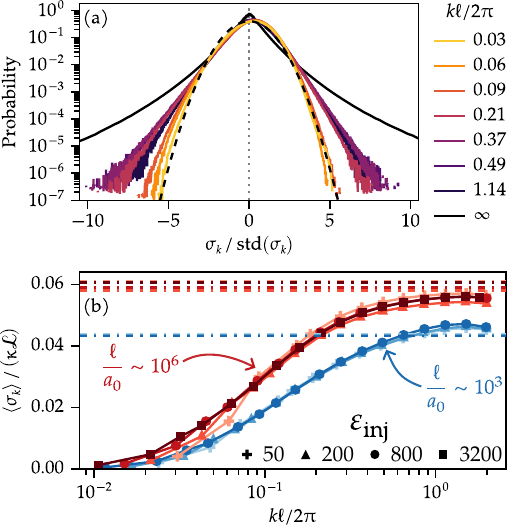}
  \end{center}
  \caption{\label{fig:line_production}%
      Local vortex line production rate $\sigma_k = \svec' \cdot
      \grad\vs^{<k} \cdot \svec'$ associated to the coarse-grained self-induced velocity
      $\vs^{<k}$.
      (a)~Probability density function (PDF) of
      $\sigma_k$, obtained for different filtering scales $k$ (run $\epsInj =
      3200$, $\azero = 10^{-7}$).
      A Gaussian PDF (dashed line) is shown for reference.
      (b) Mean value of $\sigma_k$ from different runs.
      Horizontal dash-dotted lines correspond to $\Csigma = \epsInj / (\kappa \Lcal^2 \tau_\ell)$
      where $\tau_\ell \approx (\kappa^2 / 4\pi) \ln(\ell / a_0)$ is the vortex tension.
      Two sets of curves are shown with $\ell / \azero \sim 10^{6}$ (red) and $10^{3}$ (blue),
      see the End Matter (Table~\ref{tab:simulations}) for precise values.
      Different markers correspond to different energy injection rates $\epsInj$.
      Statistics are accumulated over a large number of flow snapshots (between 48 and 99, depending on the run).
  }
\end{figure}

  To confirm the robustness of these alignments,
  in Fig.~\ref{fig:line_production} we examine the statistical properties of the local
  vortex production rate $\sigma_k$.
  The probability distribution functions (PDFs) of $\sigma_k$, shown in Fig.~\ref{fig:line_production}(a)
  for a single run, display clear deviations from a Gaussian, being negatively skewed and
  presenting increasingly heavier tails as smaller-scale motions are taken into account.
  Furthermore, as detailed in Fig.~\ref{fig:line_production}(b) for all different runs, the mean values of
  $\sigma_k$ (linked to nonlocal transfers) are positive and increase as the quantum scales are approached.
  When normalized by the effective vorticity $\omega_\ell \equiv \kappa \Lcal$,
  curves collapse for fixed $\ell / a_0$, reaching a plateau near $k = k_\ell$ at a value $\Csigma$ between 0.04 and 0.06.
  However, note that these values can depend on the finite
  numerical resolution of vortex filaments, which suppresses high-curvature regions
  likely leading to an overestimation of the alignments~\cite{Supplemental}.

  We can link the observed plateaus in Fig.~\ref{fig:line_production}(b) to energy transfers from inertial to quantum scales.
  Indeed, assuming that the identified line production mechanism is the sole responsible for such energy
  transfers, we can write the large-scale
  energy budget as ${\dd \Ecal_{k_\ell}} / {\dd t} = \epsInj - \epsSigma = 0$
  in the steady state, where $\epsSigma = \Lcal \Lmean{\sigma_{k_\ell}} \tau_\ell$.
  Here $\tau_\ell \approx ({\kappa^2} / {4\pi}) \ln(\ell / a_0)$
  is the quantum-scale energy per unit vortex length or vortex tension~\cite{Barenghi2023}.
  Setting $\Lmean{\sigma_{k_\ell}} = \Csigma \kappa \Lcal$, the above energy budget
  results in $\Csigma = \epsInj / (\kappa \Lcal^2 \tau_\ell)$.
  This energy-based estimate of $\Csigma$ [dash-dotted lines in
  Fig.~\ref{fig:line_production}(b)] closely agrees with the actual plateau
  values, confirming the validity of the assumed large-scale energy budget and thus the
  dominant role of the identified energy transfer mechanism.
  Moreover, since this energy is ultimately dissipated at quantum scales,
  one can now equate $\epsSigma$ to the energy dissipation rate, traditionally written $\varepsilon = \nu' \omega_\ell^2$
  where $\nu' = \zeta \kappa$ is an effective kinematic viscosity~\cite{Vinen2002,Bradley2006},
  which leads to $\zeta = (\Csigma / 4\pi) \ln(\ell / a_0)$.
  For $\ell / a_0 \sim 10^{6}$ we find $\zeta \approx 0.065$.
  While this value is likely overestimated as
  discussed above, it stays comparable to past \HeII{} measurements in decaying quasi-classical QT near absolute zero,
  which have reported $\zeta \approx 0.003$~\cite{Walmsley2007,Walmsley2014} from impulsive spin-down experiments and $\zeta \approx 0.08$
  from ion-jet and grid-generated QT (although with a possible effect of the container boundaries)~\cite{Zmeev2015}.

\begin{figure}
  \begin{center}
    \includegraphics[width=\columnwidth]{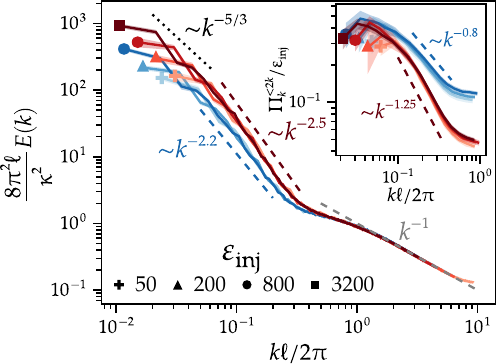}
  \end{center}
  \caption{\label{fig:energy_spectra}%
    Kinetic energy spectrum from different runs.
    Two sets of curves are shown with $\ell / \azero \sim 10^{6}$ (red) and $10^{3}$ (blue),
    see the End Matter (Table~\ref{tab:simulations}) for precise values.
    Different markers correspond to different energy injection rates $\epsInj$.
    The spectra are normalized according the analytical prediction $E(k) = (\kappa^2 / 4\pi \ell^2) k^{-1}$ in the quantum scales.
    Inset: Local energy flux $\Pi_k^{<2k}$ for the different runs.
    Note the different horizontal axes in the main panel and the inset.
    Spectra are averaged over several flow snapshots within the statistically steady state (see Table~\ref{tab:simulations}).
    Light colored bands indicate 95\% confidence intervals.
  }
\end{figure}

The existence of nonlocal energy transfers is expected to lead to the breakdown of
K41 scaling laws in QT.
This is confirmed in Fig.~\ref{fig:energy_spectra}, which shows the kinetic
energy spectra obtained from all numerical runs.
The normalized spectra collapse at quantum scales, meaning that they all follow
the analytical prediction $E(k) = (\kappa^2 \Lcal / 4\pi) k^{-1}$
associated to an isolated straight vortex for $k\azero \ll 1$~\cite{Araki2002,Polanco2025b}.
Note that a production of vortex lines directly increases the energy content
throughout the quantum scales, shifting the spectrum upwards.
At inertial scales
($k \ll k_\ell$)
the spectra become much steeper, indicating the emergence of
collective dynamics and energy-containing vortex bundles generally
associated to a partial polarization of the tangle.
  In fact, within the limited extent of scales allowed by our simulations, the
  spectrum is significantly steeper than Kolmogorov's $k^{-5/3}$ law over scales
  $k \gtrsim k_\ell/20$, as confirmed by compensated spectra~\cite{Supplemental}.
  This suggests
a higher degree of polarization
than that traditionally associated to quasi-classical QT~\cite{Lvov2007,Polanco2021}.
As expected from Eq.~\eqref{eq:energy_transfer_nonlocal}, the deviations weaken
as the ratio $\ell / \azero$ is decreased from $\sim 10^6$ (red curves) to
$\sim 10^3$ (blue curves).

Despite these major deviations, one can attempt to reconcile the observed energy
spectra with K41 theory by considering a nonconstant local energy flux
which, for simplicity, behaves as $\varepsilon_{\text{local}}(k) \sim k^{-\alpha}$.
Then, dimensional analysis leads to $E(k) \sim \varepsilon_{\text{local}}(k)^{2/3} k^{-5/3} \sim k^{-(5 + 2\alpha)/3}$.
Taking $\Pi_k^{<2k}$ (inset of Fig.~\ref{fig:energy_spectra}) as a proxy for $\varepsilon_{\text{local}}(k)$, we find
$\alpha \approx 1.25$ and $0.8$ for the two considered values of $\ell / \azero$
over about half a decade in $k$.
These values exactly correspond to the $E(k) \sim k^{-2.5}$ and $k^{-2.2}$
fits shown
in the main panel of Fig.~\ref{fig:energy_spectra}
  (which, given the limited range of scales, should not be interpreted as actual power law scalings).
This gives strong support to the idea that, similarly to classical turbulence, inertial range scalings are
independent of the specific energy injection mechanisms and of quantum scale
dynamics, despite the lack of a constant local energy cascade.


\paragraph{Discussion}%
We have identified a novel mechanism transferring energy from inertial to quantum scales in zero-temperature QT.
Besides resulting in steepening of the energy spectra at scales considerably larger than $\ell$, this
mechanism seems to predominantly explain how energy arrives to quantum scales
before being ultimately dissipated by phonon emission,
  thus avoiding the proposed bottleneck picture which would result if such transfers were
  uniquely driven by a cascade of Kelvin waves~\cite{Lvov2007,Barenghi2014}.
The crucial realization enabling these transfers is that, while individual
quantum vortices cannot be stretched in the classical
sense~\cite{Alamri2008,Barenghi2023}, nothing disallows them from being
elongated by large-scale velocity gradients, as vortex length is not a
conserved quantity.
Importantly, this mechanism transfers energy from large scales to \emph{all} quantum scales at
once, which highlights the relevance of the scale separation between $\ell$
and $\azero$ in this process.

One major remaining question is whether nonlocal transfers vanish at larger scales as the
extent of the inertial range (or the Reynolds number) is increased.
Our simulations suggest that this is indeed the case,
as the average production of vortex line approaches zero as larger scales are reached [Fig.~\ref{fig:line_production}(b)].
Physically, this may be explained by an increasing decorrelation between the
orientations of individual quantum vortices and of the resulting coarse-grained
vorticity at large scales.
Further extending the inertial range is currently out of reach for
existing numerical approaches and requires further methodological breakthroughs.
Besides, the alignment between velocity gradients, strain rate eigenvectors, and vorticity at different scales will
be the subject of a future investigation.

Being based on an alignment between large-scale velocity gradients and quantum vortices, we
expect the identified mechanism to
  be relevant in all forms of quasi-classical QT (where
  vortices self-organize into polarized bundles) independently of the turbulence
  generation mechanism.
  In particular, we have numerically verified that this is the case in
  decaying settings at zero temperature~\cite{Supplemental}.
  We also expect this mechanism to be a major feature of finite-temperature QT,
where vortices coexist and interact
with a viscous normal fluid via a mutual friction force,
and where similar alignments are likely.
The two components are expected to be tightly coupled at large scales, roughly exhibiting one-fluid behavior.
The identified nonlocal transfers can be expected to promote their decoupling
and thus the mutual friction between the two components, potentially leading to temperature-dependent
energy spectra as the quantum scale is approached.

While our work is motivated by isotropic turbulence in \HeII{}, the identified mechanism
may also be relevant in rotating settings or in other superfluids where the
ratio $\ell / \azero$ also spans several orders of magnitude in realistic settings.
This can be the case in ${}^3$He-B, where $\azero \sim 10^{-8}\,\text{m}$ while
values $\ell \sim 10^{-4}\,\text{m}$ can be experimentally achieved~\cite{Bradley2006}.
The separation of scales can be even more dramatic in the core of neutron stars~\cite{Lattimer2004}, where
$\azero \sim 10^{-14}\,\text{m}$~\cite{Seveso2016,Liu2025} 
while the inter-vortex distance associated to the star rotation alone can be estimated as
$\ell \sim 10^{-5}\,\text{m}$~\cite{Graber2017}.

\paragraph{Acknowledgments}We acknowledge fruitful discussions with L. Galantucci, M. Gibert, C. Bret and P. Diribarne.
This work was supported by the French ANR through the project ANR-23-CE30-0024-04 (QuantumVIW).
This project was granted access to computational resources of CINES and IDRIS under the allocation 2025-A0192A00611 made by GENCI.
Computations were also performed using the GRICAD infrastructure which is supported by the Grenoble research communities.

\bibliography{biblio}


\bigskip
\onecolumngrid
\begin{center}
  \textbf{\large End Matter}
\end{center}
\twocolumngrid

\paragraph{Definition of forcing velocity}%
In analogy with Eq.~\eqref{eq:energy_flux}, one can express the cumulative energy injection rate up to scale
$k$ due to an external velocity $\vf$ as
\begin{equation}
  \Fcal_k
  \equiv \left. \frac{\dd \Ecal_k}{\dd t} \right|_{\vf}
  = \frac{\kappa}{V} \Cint \left( \svec' \times \vs^{<k} \right) \cdot \vf \, \dd \xi.
  \label{eq:forcing_injection_rate}
\end{equation}
Clearly, choosing $\vf^0 = \alpha \svec' \times \vs^{<\kf}$ (with $\alpha > 0$ a nondimensional forcing coefficient)
ensures energy injection up to scale $k = \kf$.
  Note that a similar choice has been used in forced simulations of 2D point-vortex dynamics~\cite{Siggia1981}.
However, to avoid the unwanted injection of energy at quantum scales [see
Eq.~\eqref{eq:energy_transfer_nonlocal}], we also require $\vf$ to be
orthogonal to $\svec''$.
This leads to
\begin{equation}
  \vf(\svec)
  = \vf^0 - \frac{\vf^0 \cdot \svec''}{|\svec''|^2} \svec''
  = \alpha \frac{\svec'' \cdot \vs^{<\kf}}{|\svec''|^2} \svec' \times \svec''.
  \label{eq:forcing_velocity}
\end{equation}
From Eq.~\eqref{eq:forcing_injection_rate}, the large-scale energy injection rate associated to this choice is then
\begin{equation}
  \epsInj \equiv \Fcal_{\kf} = \alpha \frac{\kappa}{V} \Cint \frac{\left| \svec'' \cdot \vvec^{<\kf} \right|^2}{|\svec''|^2} \, \dd\xi.
\end{equation}
In practice, we dynamically tune the value of $\alpha$ to achieve a constant
energy injection rate $\epsInj$, similarly in spirit to commonly used forcing schemes in
classical turbulence~\cite{Lamorgese2005}.

\paragraph{Fast evaluation of Biot--Savart integrals}%
In the following, we summarize the FFT-based approach used to efficiently evaluate
Biot--Savart integrals in periodic domains.
For details the reader is referred to Ref.~\cite{Polanco2025b}.
  Our approach is inspired by fast Ewald summation methods~\cite{Arnold2005,Arnold2013}, commonly used in
  molecular dynamics to evaluate electrostatic interactions
  in periodic systems.
  Unlike other fast approaches~\cite{Baggaley2012f,Yui2025}, this method naturally accounts for the infinity of periodic
  vortex images (instead of effectively introducing a cut-off distance
  $r_{\text{cut}} \sim \Lbox$~\cite{Hanninen2014}), while also providing highly accurate
  Fourier-space velocity fields at no extra cost, facilitating the evaluation of energy spectra and
  fluxes~\eqref{eq:energy_flux}.
  The methods are implemented in the open-source GPU-accelerated VortexPasta.jl
  solver~\cite{Polanco2025c}.

Similarly to standard Ewald summation methods, the main idea is to
split the self-induced velocity field~\eqref{eq:BiotSavart} as
$\vs(\xvec) = \vs\near(\xvec) + \vs\far(\xvec)$, where $\vs\near$ and $\vs\far$
respectively account for the influence of vortex elements located near and far from the
evaluation point $\xvec$.
To obtain expressions for these two components, we first introduce the vector
potential $\psisvec(\xvec)$ such that $\vs = \curl \psisvec$.
It is related to the vorticity by Poisson's equation $\vortsvec = \curl \vs = -\laplacian \psisvec$,
whose solution can be written as the convolution $\psisvec = G \ast \vortsvec$
where $G(\rvec) = 1 / (4\pi r)$ is the Green's function associated to the 3D Poisson's equation.
In the context of the VFM, this becomes
\begin{equation}
  \psisvec(\xvec)
  = \frac{\kappa}{4\pi} \Cint \frac{\dd\svec}{|\svec - \xvec|}.
  \label{eq:vector_potential}
\end{equation}
It can be readily shown that $\curl \psisvec$ leads to the Biot--Savart law~\eqref{eq:BiotSavart}.
As a side note, knowing $\psisvec$ at vortex locations $\svec \in \Ccal$ conveniently
allows to evaluate the total kinetic energy per unit mass as $E = \frac{\kappa}{2V} \Cint \psisvec(\svec) \cdot \dd\svec$.

The main idea of standard Ewald methods is to split the Green's function $G(\rvec)$ as
\begin{equation}
  G(\rvec)
  = \frac{\erfc(\alpha r)}{4\pi r} + \frac{\erf(\alpha r)}{4\pi r}
  = G\near(\rvec) + G\far(\rvec),
  \label{eq:vector_potential_split}
\end{equation}
where $\erf(x)$ is the error function and $\erfc(x) = 1 - \erf(x)$.
Here $\alpha$ is a tunable splitting parameter (an inverse length scale) setting the
transition between near- and far-ranged interactions.
Taking the gradient of Eq.~\eqref{eq:vector_potential_split}, a similar splitting can then be obtained for the Biot--Savart kernel,
$\grad G(\rvec) = -\rvec / (4\pi r^3) = \grad G\near(\rvec) + \grad G\far(\rvec)$,
with
\begin{align}
  \grad G\near(\rvec) &= - \left[ \erfc(\alpha r) + \frac{2\alpha r}{\sqrt{\pi}} e^{-(\alpha r)^2} \right] \frac{\rvec}{4\pi r^3}, \\
  \grad G\far(\rvec) &= - \left[ \erf(\alpha r) - \frac{2\alpha r}{\sqrt{\pi}} e^{-(\alpha r)^2} \right] \frac{\rvec}{4\pi r^3},
\end{align}
allowing to define
$\vs\near = \grad G\near \ast \vortsvec$ and $\vs\far = \grad G\far \ast \vortsvec$.
With the above choice of splitting, it can be shown that $\vs\far$ is nothing else than the velocity field induced by a Gaussian-filtered vorticity field
$\vortsvec\far(\xvec) = (\varphi_\alpha \ast \vortsvec)(\xvec)$
with $\varphi_\alpha(\rvec) = (\alpha / \sqrt{\pi})^3 e^{-(\alpha r)^2}$.

The first crucial point here is that, unlike the full Biot--Savart kernel, $\grad G\near(\rvec)$ decays
very quickly with $r$, which means that one only needs to consider nearby pairs
within a radius $\rcut = \beta / \alpha$ to obtain an arbitrarily accurate estimation of
$\vs\near$.
Here $\beta$ is a nondimensional parameter controlling truncation
errors~\cite{Polanco2025b}, which we set to $\beta = 3.5$ for $10^{-6}$
relative accuracy.
Secondly, $\grad G\far(\rvec)$ is nonsingular and well-behaved near $r = 0$, unlike $\grad G(\rvec)$.
In periodic domains, this allows to represent $\vs\far$ as a truncated Fourier
series up to wavenumber $\kmax = 2\beta \alpha$~\cite{Polanco2025b}.
Then, the Fourier coefficients of $\vs\far$ are simply
\begin{equation}
  \vshat\far(\kvec)
  = \frac{i \kvec \times \vortsshat\far(\kvec)}{|\kvec|^2}
  = \frac{i \kvec \times \vortsshat(\kvec)}{|\kvec|^2} \, e^{-k^2 / 4\alpha^2},
  \label{eq:velocity_far_fourier}
\end{equation}
where $\vortsshat(\kvec) = \frac{\kappa}{V} \Cint e^{-i \kvec \cdot \svec} \, \dd \svec$
is a Fourier coefficient of the superfluid vorticity field.

In practice, to obtain $\vs\far$, we start by approximating the $\vortsshat(\kvec)$ line integral using 3-point
Gauss--Legendre quadratures on each discrete vortex segment, requiring $3N$
spline interpolations of $\svec(\xi)$ and $\svec'(\xi)$.
This results in a Fourier sum over $3N$ nonequispaced points, which is then
efficiently evaluated using the type-1 nonuniform FFT (NUFFT)
algorithm~\cite{Dutt1993,Greengard2004}.
This step yields $\vortsshat(\kvec)$ coefficients on a uniform 3D Fourier grid
truncated at $\pm\kmax$ in each direction.
Then, once Eq.~\eqref{eq:velocity_far_fourier} has been evaluated, we
interpolate $\vs\far$ at vortex locations in physical space using the type-2 NUFFT.
Overall, by properly choosing the splitting parameter $\alpha$, this method
leads to $O(N \log N)$ complexity, enabling the fast and arbitrarily accurate
evaluation of Biot--Savart interactions in dense vortex systems.

\paragraph{Spatial discretization and timestepping}%
Vortex lines are discretized as quintic splines, ensuring an accurate
estimation and interpolation of geometrical quantities such as $\svec$, $\svec'$ and
$\svec''$.
All line integrals are evaluated using 3-point Gauss--Legendre quadratures between
neighboring vortex points, and the relative accuracy of Biot--Savart
computations is kept below $10^{-6}$~\cite{Polanco2025b}.
As in previous works~\cite{Araki2002,Adachi2010,Baggaley2012,Galantucci2019}, the distance
$\Delta\xi$ between vortex points is kept within $[\deltaRes, 2\deltaRes)$
using adaptive line refinement.
The resolution is set to $\deltaRes \approx \ell/6$, adjusted according to
the mean inter-vortex distance $\ell$ in each simulation.
  We have checked that doubling the resolution to $\deltaRes \approx \ell/12$ does not change the overall conclusions
  of our work~\cite{Supplemental}.
Following standard practice, filaments are reconnected when their distance is below $\deltaRes$, but only if
this leads to loss of vortex length (effectively dissipating energy at quantum
scales) and if the reconnecting segments are not locally parallel~\cite{Baggaley2012a}.
Small vortex rings with diameters below about $2\deltaRes$ are also removed from the simulations.
Equation~\eqref{eq:ds_dt} is advanced in time using an adaptive 4th-order Runge--Kutta scheme,
  with a time step kept below $\Delta t_{\text{KW}} = 2 (\deltaRes^2 / \kappa) \left[ \ln(\deltaRes / \pi a_0) + (1/2) - (\Delta + \gamma) \right]^{-1}$
  where $\gamma$ is the Euler--Mascheroni constant.
  This quadratic stability criterion is related to the period of Kelvin wave oscillations at the scale $\deltaRes$.

\paragraph{Numerical runs}%
In Table~\ref{tab:simulations} we detail relevant parameters associated to the VFM simulations discussed in the main text.

\paragraph{Estimation of confidence intervals}%
Confidence intervals for time-averaged quantities in Figs.~\ref{fig:fluxes3200}
and~\ref{fig:energy_spectra} are based on the central limit theorem.
Assuming $n = \lfloor \Tsim / \Tint \rfloor$ statistically independent
snapshots (see Table~\ref{tab:simulations} for definitions and values),
the statistical uncertainty is written as $\mu \pm c \frac{\sigma}{\sqrt{n}}$
where $\mu$ is the estimated mean and $\sigma$ the unbiased sample standard deviation.
To obtain 95\% confidence intervals and to account for the finite sample size,
$c$ is chosen as the 97.5th percentile of Student's $t$ distribution with $n - 1$ degrees of freedom.
In practice, for the estimation of confidence intervals we take $n = \min\left\{ \lfloor \Tsim / \Tint \rfloor, n_\text{snap} \right\}$
where $n_{\text{snap}} = \Tsim / \dtSnap + 1$ is the number of considered flow snapshots.

\begin{widetext}

\begin{table}[tb]
  \caption{\label{tab:simulations}%
    Physical and numerical parameters of the simulations.
    The superfluid Reynolds number is $\ReyKappa = \vrms \Lint / \kappa$~\cite{Finne2003,Salort2011,Polanco2025a,Bret2025}
    where $\vrms^2 = (2/3) \Ecal_{\kell}$ is the large-scale velocity variance (associated to turbulent fluctuations).
    The integral scale is estimated as $\Lint = \frac{\pi}{2 \vrms^2} \int_0^{\kell} k^{-1} E(k) \, \dd k$
    where $E(k)$ is the energy spectrum.
    $N$ is the number of discrete vortex points.
    $\Tsim/\Tint$ is the simulation time in units of the eddy turnover
    (or flow correlation) time $\Tint = \Lint / \vrms$.
    All results reported in this Letter are time-averaged within the statistically-steady state $t \in [0, \Tsim]$.
    In particular, $\dtSnap / \Tint$ is the time interval (in correlation
    time units) between flow snapshots used to obtain energy spectra and fluxes.
    See main text for other definitions.
  }
  \begin{ruledtabular}
    \begin{tabular}[c]{ccccccccccc}
      Run & $\epsInj$ & $\azero$ & \ReyKappa & $\Lint$ & $\Lint/\ell$ & $\ell/\azero$ & $\ell / \deltaRes$ & $N$ & $\Tsim/\Tint$ & $\Delta t_{\text{snap}} / \Tint$ \\
      \hline
      A1 & $50$   & $10^{-7}$ &  $5.2$ & $1.23$ &  $6.3$ & $1.9 \times 10^{6}$ & $5.89$ & $1.3 \times 10^{5}$ & $18.9$ & 1.72  \\
      A2 & $200$  & $10^{-7}$ &  $8.9$ & $1.26$ &  $9.4$ & $1.3 \times 10^{6}$ & $6.55$ & $4.5 \times 10^{5}$ &  $8.4$ & 0.56  \\
      A3 & $800$  & $10^{-7}$ & $14.5$ & $1.24$ & $13.1$ & $9.5 \times 10^{5}$ & $6.18$ & $1.2 \times 10^{6}$ & $37.7$ & 4.71  \\
      A4 & $3200$ & $10^{-7}$ & $23.1$ & $1.23$ & $18.3$ & $6.7 \times 10^{5}$ & $6.24$ & $3.4 \times 10^{6}$ & $15.3$ & 0.51  \\
      \hline
      B1 & $50$   & $10^{-4}$ &  $5.4$ & $1.14$ &  $7.6$ & $1.5 \times 10^{3}$ & $6.37$ & $3.1 \times 10^{5}$ & $13.2$ & 0.41 \\
      B2 & $200$  & $10^{-4}$ &  $8.2$ & $1.10$ & $10.5$ & $1.0 \times 10^{3}$ & $6.47$ & $9.3 \times 10^{5}$ & $12.9$ & 0.68 \\
      B3 & $800$  & $10^{-4}$ & $13.8$ & $1.14$ & $15.7$ & $7.3 \times 10^{2}$ & $6.11$ & $2.6 \times 10^{6}$ &  $7.4$ & 0.53 \\
    \end{tabular}
  \end{ruledtabular}
\end{table}

\end{widetext}


\onecolumngrid
\clearpage

\begin{center}
  \textbf{\large Supplemental Material for \enquote{Nonlocal Energy Transfer Mechanism in Three-dimensional Quantum Turbulence}}
\end{center}

\setcounter{equation}{0}
\setcounter{figure}{0}
\setcounter{table}{0}
\setcounter{page}{1}

\renewcommand{\thesection}{SM\arabic{section}}
\renewcommand{\thetable}{SM\arabic{table}}
\renewcommand{\thefigure}{SM\arabic{figure}}
\renewcommand{\theequation}{SM\arabic{equation}}

\newcommand*{\bhat}{\hat{\bm{b}}}  
\newcommand*{\vsQuantum}{\vs^{>k_\ell}}
\newcommand*{\curv}{|\svec''|}
\newcommand*{\deltaRec}{\delta_{\text{rec}}}

\section{Localized induction approximation and velocity at quantum scales}\label{sec:curvature_lia_vsmall}

In the main text, the localized induction approximation (LIA) is used to obtain
an estimation of the fine-grained velocity $\vsQuantum$, which then leads to
an analytical expression for the nonlocal energy flux from large to quantum scales.
Using Eq.~(2) of the main text, the LIA-based estimation can be expressed as
\begin{equation}
  \vsQuantum(\svec) \approx \vs^{\ell}(\svec)
  = C_\ell \, |\svec''| \, \bhat(\svec)
  \quad \text{with} \quad C_\ell = \frac{\kappa}{4\pi} \ln\left( \frac{2\ell}{e^\Delta a_0} \right)
  \quad \text{and} \quad \bhat(\svec) = \frac{\svec' \times \svec''}{|\svec''|},
  \label{eq:LIA_vsmall}
\end{equation}
where $\bhat$ is the local (unitary) binormal vector.

In the following we numerically evaluate the relevance of
Eq.~\eqref{eq:LIA_vsmall} as an approximation of the actual quantum scale velocity $\vsQuantum$.
We focus in particular on the accuracy of this estimate as a function of the
local curvature $\svec''$.
Indeed, as stated in Eq.~(2) of the main text, the above approximation can be expected to fail
in high curvature regions ($\curv \ell \gg 1$) -- generally associated to cusps generated before and after
a reconnection event~\cite{deWaele1994,Kivotides2001a,Boue2013a,Hanninen2013,Villois2017}.
On the other hand, Eq.~\eqref{eq:LIA_vsmall} also requires nonlocal effects to be negligible.
However, in spatial regions where vortices are more densely packed than
average (i.e.\ the local vortex distance is smaller than $\ell$), the actual
$\vsQuantum$ may include a contribution from neighboring vortices which is not
accounted for by the LIA.
Such nonlocal contribution may even be expected to be dominant where the LIA-based velocity is small,
i.e.\ in small curvature regions ($\curv \ell \ll 1$).
These hypotheses are confirmed numerically below.

\begin{figure}[b]
  \begin{center}
    \includegraphics[width=\textwidth]{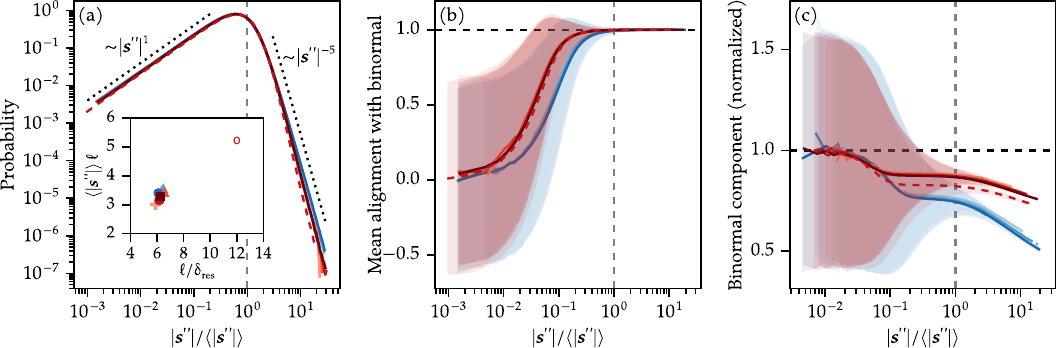}
  \end{center}
  \caption{%
    Comparison between quantum-scale velocity $\vsQuantum$ and its LIA-based approximation [Eq.~\eqref{eq:LIA_vsmall}]
    from different runs.
    (a) Curvature PDFs standardized by the mean curvature $\Lmean{\curv}$.
    Power law scalings (dotted lines) are plotted for reference.
    The inset shows the normalized mean curvature $\Lmean{\curv} \ell$ as a
    function of the numerical line resolution $\ell / \deltaRes$.
    (b) Mean alignment $\Lmean{( \vsQuantum \cdot \bhat ) / |\vsQuantum|}$ between the binormal
    vector $\bhat$ and quantum-scale velocity $\vsQuantum$, conditioned on local curvature values.
    (c) Mean value $\Lmean{ \vsQuantum \cdot \bhat }$ of the binormal component of the quantum-scale velocity,
    normalized by the LIA-based value $C_\ell \curv$.
    In (b) and (c), light colored bands around each curve have a height
    $2\sigma$ where $\sigma$ is the standard deviation of the respective
    distributions.
    Colors are the same as in the main text (see e.g.~Fig. 4).
    In each plot, dashed red lines correspond to an extra run with double
    resolution ($\deltaRes \approx \ell / 12$, $a_0 = 10^{-7}$ and $\epsInj = 800$),
    detailed in Sec.~\ref{sub:resolution_effects}.
  }\label{fig:curvature_lia_vsmall}
\end{figure}

Before evaluating the impact of the local curvature on the relevance of the LIA-based approximation,
it is instructive to consider the distribution of the filament curvature itself.
For this, we plot in Fig.~\ref{fig:curvature_lia_vsmall}(a) the PDF of the
curvature $\curv / \Lmean{\curv}$ from all runs considered in the main text (solid lines; Table I of End Matter),
which are performed at the same normalized line resolution $\deltaRes \approx \ell / 6$.
Remarkably the PDFs showcase self-similar behavior, collapsing independently of the value of $\epsInj$ or $a_0$.
For small curvatures, the PDFs display a clear linear behavior which has
already been observed in quantum turbulence simulations of the
VFM~\cite{Kondaurova2014} (at finite temperatures) and of the
Gross--Pitaevskii~\cite{Villois2016a} models.
On the other hand, the PDFs at large curvatures roughly decay following power
laws (with slight variations between runs) which differ from the exponential
decays observed in the cited works.
In the case of Ref.~\cite{Kondaurova2014}, this is likely explained by the
smoothing of small-scale line fluctuations due to mutual friction at finite temperatures.
Note, however, that the actual curvature values are affected not only by the physical scale $\ell$
but also by the numerical resolution.
Indeed, reducing the distance $\deltaRes$ between discrete points leads to larger curvatures,
as we verify from a separate run at $\deltaRes = \ell / 12$ (detailed in Sec.~\ref{sub:resolution_effects}).
While the resulting standardized PDF (dashed red line) agrees well with the other runs, the mean curvature
increases when normalized with $\ell$ [inset of Fig.~\ref{fig:curvature_lia_vsmall}(a)].
This illustrates the limitations of the VFM at describing the small scale geometry of quantum vortices,
and the need for resolutions $\deltaRes \ll \ell$ to ensure that
turbulent scales are not affected.

To assess the relevance of the LIA-based approximation, we start by numerically
evaluating the relative orientation between $\vsQuantum$ and the binormal direction
$\bhat$.
In Fig.~\ref{fig:curvature_lia_vsmall}(b) we plot their average alignment $\Lmean{( \vsQuantum \cdot \bhat ) / |\vsQuantum|}$,
expected to be 1 according to Eq.~\eqref{eq:LIA_vsmall}.
For large curvatures, the quantum-scale velocity is perfectly aligned with
$\bhat$, suggesting that it is predominantly determined by local contributions
while effects of neighboring vortices are negligible.
Interestingly, the opposite occurs in the small curvature limit, where the
average alignment vanishes and its standard deviation (represented by colored
bands around each curve) approaches the $\sigma = 1 / \sqrt{3}$ value
corresponding to a uniform distribution.
In other words, in this limit the orientation of $\vsQuantum$ is uncorrelated
with $\bhat$ and thus the LIA-based estimation fails, very likely due to the
dominance of nonlocal effects as suggested earlier.
That said, in the core of the distribution ($\curv \sim \Lmean{\curv}$) the alignment is
very close to 1, suggesting that $\vsQuantum$ is well estimated over a dominant fraction of vortex segments.

This is confirmed in Fig.~\ref{fig:curvature_lia_vsmall}(c), which considers the projection $\vsQuantum \cdot \bhat$
(equal to $|\vsQuantum|$ if the alignment is 1, i.e.\ for large curvatures).
Values are normalized by the LIA-based estimation $C_\ell \curv$.
In the core of the distribution, for $0.1 \lesssim \curv / \Lmean{\curv} \lesssim 1$, the mean projection roughly displays a plateau
at values of order 1, indicating that Eq.~\eqref{eq:LIA_vsmall} provides a
reasonable approximation of the actual quantum-scale velocity, despite minor deviations and a clear effect of $\ell / a_0$.
Interestingly, for larger curvatures where the alignment was near perfect, the
agreement in the actual velocity magnitude tends to decrease.
This can be explained by the fact that the filament geometry varies over a scale
of the order of the local curvature radius, which in this case is much smaller
than the scale $\epsilon = \ell$ used in the LIA-based expression (in other
words, the $\curv \ell \ll 1$ condition is not met).

Overall, it can be concluded that Eq.~\eqref{eq:LIA_vsmall} approximates
reasonably well the orientation and (up to an order-one constant) the magnitude
of the quantum scale velocity $\vsQuantum$.
The main exception are regions of very small curvature where the local
contribution is negligible and nonlocal interactions persist.
These regions, where filaments are nearly straight and thus cannot significantly
stretch locally, are not expected to contribute much to nonlocal energy transfers,
which will be rather governed by moderate to high-curvature vortex sections
whose local length can change more easily due to a large-scale velocity field.

\section{Nonlocal energy flux and vortex line production conditioned on local curvature}\label{sec:nonlocal_flux_conditioned_on_curvature}

In the following we evaluate the contribution of different curvature ranges to
the nonlocal energy flux and to vortex line production by the large-scale velocity.
For this, we start by writing the nonlocal energy flux as
$\Pi_k^{>k_\ell} = \kappa \Lcal \Lmean{(\svec' \times \vs^{>k_\ell}) \cdot \vs^{<k}}$,
which is equivalent to the definition given in the main text.
If we now consider the curvature PDF $P(\curv)$
[Fig.~\ref{fig:curvature_lia_vsmall}(a)], we can expand the nonlocal flux as
\begin{equation}
  \Pi_k^{>k_\ell} = \kappa \Lcal \int_0^{\infty} \Lmean{(\svec' \times \vs^{>k_\ell}) \cdot \vs^{<k} | \rho} \, P(\rho) \, \dd\rho,
  \label{eq:nonlocal_flux_curvature}
\end{equation}
where $\Lmean{X | \rho}$ denotes the conditional average of a quantity $X$ on curvature values $\curv = \rho$.
Then, the contribution of curvature $\rho$ to the global average $\Lmean{X}$ is formally $\Lmean{X | \rho} P(\rho) \dd\rho$,
which can be readily evaluated numerically once the curvature has been discretized into bins $[\rho_i, \rho_{i + 1})$.

\begin{figure}
  \begin{center}
    \includegraphics[width=\textwidth]{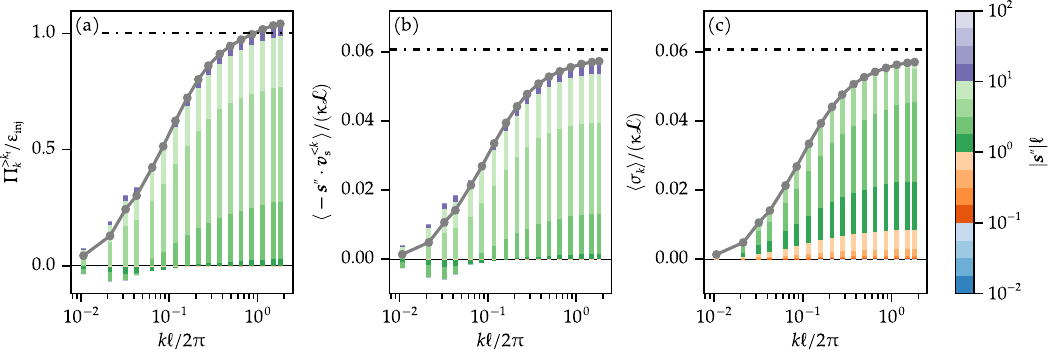}
  \end{center}
  \caption{%
    Contribution of different curvature ranges to nonlocal energy fluxes and vortex line production.
    (a) Nonlocal energy flux $\Pi_k^{k_\ell} \equiv \kappa \Lcal \Lmean{(\svec' \times \vs^{>k_\ell}) \cdot \vs^{<k}}$;
    (b) average anti-alignment between curvature $\svec''$ and large-scale velocity $\vs^{<k}$;
    (c) average vortex line production $\sigma_k = \svec' \cdot \grad\vs^{<k} \cdot \svec'$.
    In (b) and (c), horizontal dashed lines represent $C_\sigma = \epsInj / (\kappa \Lcal^2 \tau_\ell)$ defined in the main text.
    Colors represent different curvature ranges (see colorbar).
    The unconditioned means are represented by gray circles and lines.
    Results obtained from run A4 ($\epsInj = 3200$, $\azero = 10^{-7}$, $\deltaRes \approx \ell/6$) and accumulated across 99 flow snapshots.
  }\label{fig:curvature_nonlocal_flux}
\end{figure}

\begin{figure}
  \begin{center}
    \includegraphics[width=\textwidth]{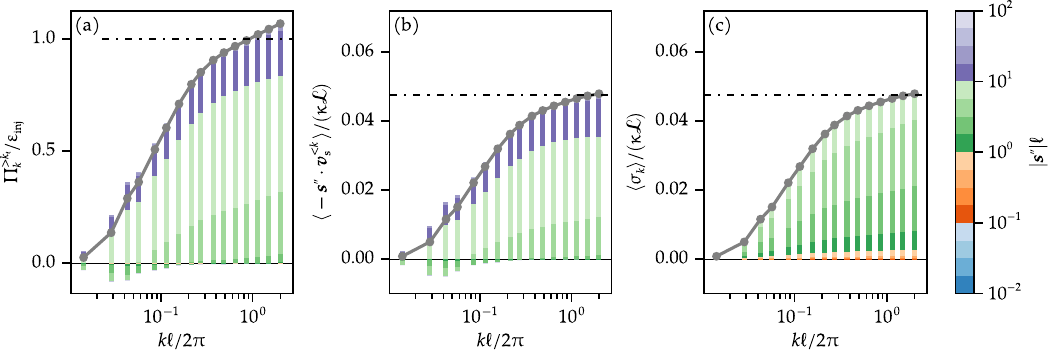}
  \end{center}
  \caption{%
    Same as Fig.~\ref{fig:curvature_nonlocal_flux} for the extra high-resolution run
    ($\epsInj = 800$, $\azero = 10^{-7}$, $\deltaRes \approx \ell/12$).
  }\label{fig:curvature_nonlocal_flux_highres}
\end{figure}

In Fig.~\ref{fig:curvature_nonlocal_flux}(a), we present the contribution of
different curvature ranges to the nonlocal energy
flux from one of our runs.
For all filtering scales, it is clear that it is mainly the large curvatures
$\curv > 1/\ell$ which contribute to the integral in
Eq.~\eqref{eq:nonlocal_flux_curvature}.
Interestingly, at small wavenumbers, curvatures $\curv \gtrsim 1 / \ell$ can carry a negative contribution to the integral.
As discussed in the main text [Eq.~(5)], using the LIA to approximate
$\vs^{>k_\ell}$ leads to a proportionality relation between the nonlocal flux $\Pi_k^{>k_\ell}$
and the mean vortex line density production rate $\Scal_k = \Lcal \Lmean{-\svec'' \cdot \vs^{<k}}$.
In Fig.~\ref{fig:curvature_nonlocal_flux}(b) we verify that the contributions
to $\Pi_k^{>k_\ell}$ from different curvature ranges directly translate into equivalent contributions
to $\Lmean{-\svec'' \cdot \vs^{<k}}$, highlighting once again the relevance of
the LIA-based estimate and of the derived curvature-based expression for the nonlocal flux.
Note that, since they concern high curvature regions, these results may be affected by the
finite numerical resolution, which limits the range of curvatures that can be
captured numerically (as quantified in Sec.~\ref{sec:curvature_lia_vsmall}).
This is confirmed in Fig.~\ref{fig:curvature_nonlocal_flux_highres},
which shows that at higher resolutions ($\deltaRes \approx \ell/12$)
even larger curvatures can contribute to the $\Pi_k^{>k_\ell}$ line integral.

While by definition the integral of $(\svec' \times \vs^{>k_\ell}) \cdot \vs^{<k}$
quantifies an energy transfer across scales, the physical interpretation of its
local value at a given vortex location is less clear,
and the same can be said of the local anti-alignment $-\svec'' \cdot \vs^{<k}$.
Note in particular that these two quantities are not Galilean invariant, i.e.\ their
local values (and signs) depend on the chosen inertial frame of reference.
For this reason, we turn to the alternative expression of the mean vortex line density production rate,
$\Scal_k = \Lcal \Lmean{\sigma_k}$ where $\sigma_k = \svec' \cdot \grad\vs^{<k} \cdot \svec'$ [Eq.~(6) in main text].
Since $\sigma_k$ only depends on the local vortex geometry and on velocity gradients, it is Galilean invariant,
while being equivalent to $-\svec'' \cdot \vs^{<k}$ one only in an integral sense
(once integrated over a closed or infinitely periodic curve).
Moreover, only $\sigma_k$ can be directly interpreted as the linear strain along the
vortex axis due to large-scale velocity gradients, and thus as the
infinitesimal vortex line production rate per unit line.
Finally, since the quantum-scale energy is proportional to the total vortex
length via the vortex tension $\tau_\ell$, the local strain rate $\sigma_k$ has a clear interpretation in
terms of a (spatially local) energy transfer to (or from) quantum scales.

With this in mind, we now evaluate the contribution of different curvature ranges
to $\Lmean{\sigma_k}$, estimated from conditional averages $\Lmean{\sigma_k | \rho}$.
As clearly seen in Fig.~\ref{fig:curvature_nonlocal_flux}(c), this estimation
leads to a very different picture from that based either on the
$\Pi_k^{>k_\ell}$ definition [Fig.~\ref{fig:curvature_nonlocal_flux}(a)] or on
curvature anti-alignment [Fig.~\ref{fig:curvature_nonlocal_flux}(b)].
First, because high curvatures have a much weaker contribution,
while moderate curvatures $\curv \sim 1/\ell$ are those which contribute the
most to vortex line production.
Second, because all contributions are positive, meaning that coarse-grained velocity gradients
produce vortex length regardless of the filtering scale $k$ or the local vortex curvature $\curv$.
While some resolution effects persist
[Fig.~\ref{fig:curvature_nonlocal_flux_highres}(c)], the contribution of very
high curvatures ($\curv > 10/\ell$) stays negligible, which was not the case for the other estimates.
This result can be understood by considering that the orientation of vortex
filaments in high-curvature regions (typically resulting from reconnections) is
mostly random and decorrelated from the large-scale velocity gradients
$\grad\vs^{<k}$, leading on average to zero vortex line production.

\section{Steady state and time evolution of global quantities}\label{sec:time_evolution_forcing}

\begin{figure}[tb]
  \begin{center}
    \includegraphics[]{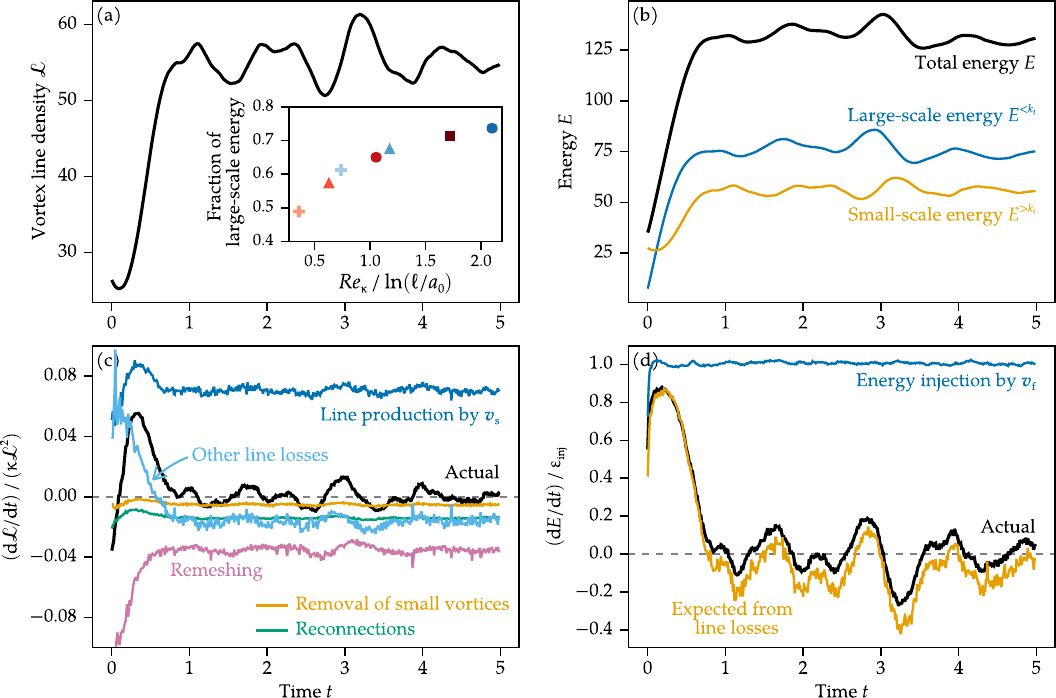}
  \end{center}
  \caption{%
    Time evolution of vortex line density and kinetic energy in a forced simulation.
    (a) Vortex line density over time.
    Inset: fraction of large-scale kinetic energy from all runs in the main text.
    (b) Small-scale, large-scale and total kinetic energy over time.
    (c) Time derivative of vortex line density (black) along with
    contributions from self-induced velocity gradients (blue),
    removal of small vortices (orange), reconnections (green), line remeshing
    (pink), and other line losses (light blue).
    (d) Time derivative of kinetic energy (black) along with the energy injection rate
    by the forcing velocity $\vf$ (blue) and expected energy balance based on dissipation by
    total line losses (orange).
    Results are obtained from a run with $\epsInj = 200$, $a_0 = 10^{-7}$ and $\deltaRes = \ell / 6$.
  }\label{fig:line_energy_budgets}
\end{figure}

In this section we provide details on the time evolution of the vortex line density
and the kinetic energy in the steady state of a forced simulation as well as in the
transient before this steady state is reached.
For this we consider a single run with energy injection rate $\epsInj = 200$,
$a_0 = 10^{-7}$ and $\deltaRes = \ell / 6$ (same parameters as Run A2 in the main
text, see Table I).
The simulation is started with 1000 randomly oriented and randomly positioned vortex rings with
radius $R = \Lbox / 6$.
As seen in Fig.~\ref{fig:line_energy_budgets}(a), the vortex line density
$\Lcal$ slightly decreases at very early times as the spatial configuration readjusts,
before steadily increasing due to the effect of the forcing velocity $\vf$.
Slightly before $t = 1$, a statistically stationary state is reached where
$\Lcal$ displays random fluctuations around an average value, as could be
expected for the steady state of a turbulent regime sustained by an external
energy injection mechanism.

The total kinetic energy [Fig.~\ref{fig:line_energy_budgets}(b)] displays
similar features to $\Lcal$, with the difference that it increases monotonously
all along the transient.
To evaluate the fraction of energy \enquote{stored} in the quantum scales
between $a_0$ and $\ell$, we decompose the kinetic energy into large- and small-scale parts.
The latter is estimated as $E^{>k_\ell} = \tau_\ell \Lcal$ where
$\tau_\ell \equiv \frac{\kappa^2}{4\pi} \ln\left( \frac{\ell}{2\pi a_0 e^{\Delta - 1}} \right)$ is
the vortex tension~\cite{Barenghi2023}, which quantifies the kinetic energy per
unit vortex length within the quantum scales.
The large-scale energy is then $E^{<k_\ell} = E - E^{>k_\ell}$.
We have verified that, estimated as such, $E^{<k_\ell}$ accurately matches the
cumulative kinetic energy $\Ecal_{k_\ell} = \int_{0}^{k_\ell} E(k) \, \dd k$
computed from the energy spectrum $E(k)$.
In the initial transient, as seen in Fig.~\ref{fig:line_energy_budgets}(b), the large-scale
energy monotonously increases due to the large-scale forcing, while the small-scale energy
displays an initial dip associated to the loss of vortex length at very early
times [Fig.~\ref{fig:line_energy_budgets}(a)].
At later times, in the statistically stationary state, both energy components
display slow fluctuations around a mean, indicating that large and small scales
are separately in equilibrium.
In this case, the small-scale energy is only slightly smaller than the energy associated to
collective vortex dynamics, illustrating the major role played by the quantum
scale in the energy balance, especially since the integral scale is relatively
small in this specific run ($\Lint / \ell \approx 9.4$).
In the inset of Fig.~\ref{fig:line_energy_budgets}(a) we report the fraction of
large-scale kinetic energy, $E^{<k_\ell} / E$, associated to each run of the main text,
showing that it slowly approaches unity as the Reynolds number $\ReyKappa$
(rescaled here by $\ln(\ell / a_0)$) is increased.

In Fig.~\ref{fig:line_energy_budgets}(c) we focus on the vortex line density budget.
Consistently with panel (a), the time derivative $\dd\Lcal/\dd t$
(black lines) fluctuates around zero in the statistically steady state.
This rate of change is expected to be
determined by the vortex line production due to the self-induced velocity field,
and by loss of vortex length associated to various numerical processes discussed in the following.
Recall that, by construction, the forcing velocity $\vf$ is orthogonal to
$\svec''$ and therefore does not directly contribute to the line budget.
The production term [blue line in Fig.~\ref{fig:line_energy_budgets}(c)] is given by $\Scal_\infty = -\frac{1}{V} \Cint \svec'' \cdot \vs \, \dd\xi$,
which stays positive and close to a value of $0.07$ when normalized by
$\kappa\Lcal^2$, quantitatively consistently with the values of $\Lmean{\sigma_k}$ at
large $k$ [Fig.~3(b) of main text].
In the steady state, this term can be expected to be balanced with line losses
associated to the finite numerical resolution $\deltaRes$ of the vortex filaments.
During our simulations, we monitor the loss of vortex length due to
reconnections (below a distance $\deltaRes$), removal of small vortex rings (radius below about $\deltaRes$),
and the remeshing algorithm, which ensures that the distance $\Delta\xi$ between vortex points stays within $[\deltaRes, 2\deltaRes)$.
Unlike the total length in Fig.~\ref{fig:line_energy_budgets}(a), these losses
are estimated by approximating filament segments as straight, thus likely
underestimating the actual losses.
Note that keeping $\Delta\xi \ge \deltaRes$ is critical to avoid a vanishingly
small time step $\Delta t \sim {\Delta\xi}_{\min}^2$ required for numerical
stability.
However, this necessarily leads to point removal and therefore to loss of vortex length.
In our simulations, we find that this procedure dominates vortex length losses,
representing about half of the total losses in the steady state [pink line in
Fig.~\ref{fig:line_energy_budgets}(c)] -- much higher than that due
to reconnections (green) and to removal of small vortex rings (orange).
Besides, note that these three processes do not fully account for the total loss of vortex length.
Indeed, other line losses not accounted for (light blue) are comparable to the losses by reconnections.
This might be explained either by the straight segment approximation mentioned
above, or perhaps by the smoothing of high curvature regions or small
wavelength fluctuations (e.g.\ associated to Kelvin wave activity).

To finish, we consider in Fig.~\ref{fig:line_energy_budgets}(d) the total kinetic energy budget.
As for $\Lcal$, the rate of change of the total kinetic energy (black line) fluctuates
around zero in the statistically steady state.
As expected, the estimated energy injection rate $\varepsilon_{\text{f}}$ by the forcing velocity $\vf$ (blue line) is
close to the prescribed energy injection rate $\epsInj$.
Assuming that dissipation occurs only at quantum scales, $\epsInj$ is expected to be balanced by the energy dissipation
associated to numerical loss of vortex length.
This dissipation can be estimated as $\varepsilon_L = -\tau_\ell \! \left.\frac{\dd \Lcal}{\dd t}\right|_{\text{losses}}$,
where $\tau_\ell$ is the vortex tension and $\left.\frac{\dd \Lcal}{\dd t}\right|_{\text{losses}}$ quantifies all sources of vortex line loss
(it is estimated as the difference between the black and blue curves in Fig.~\ref{fig:line_energy_budgets}(c)).
Using this estimation, the resulting energy budget $\dd E/\dd t =
\varepsilon_{\text{f}} - \varepsilon_L$ (orange line) is a good approximation
to the actual energy budget (black line) obtained by directly differentiating $E(t)$, confirming that energy is mostly dissipated at quantum scales
by vortex line losses.

\section{Influence of numerical parameters}

\subsection{Filament resolution}\label{sub:resolution_effects}

As already discussed, the filament resolution $\deltaRes$ can be expected to have an influence on the
results from VFM simulations, as it limits the maximum curvature $|\svec''|$
which can be captured numerically (see Fig.~\ref{fig:curvature_lia_vsmall}(a))
and ultimately the energy dissipated by loss of vortex line length.
In this section we evaluate the resulting impact of the chosen resolution on
the results discussed in the main paper.
For this, we perform an extra high-resolution run with $\deltaRes
\approx \ell/12$ (half the discretization distance used in the main text) with
an energy injection rate $\epsInj = 800$ and vortex core radius $\azero = 10^{-7}$.
The critical reconnection distance is kept equal to $\deltaRes$.
Note that, from a numerical standpoint, the cost of such simulation is at least 8 times higher than
the equivalent run at $\deltaRes \approx \ell/6$, mainly because the time step
$\Delta t$ must be divided by $\sim 4$ to ensure the numerical stability of the solver.
Statistics are accumulated over $\Tsim / \Tint \approx 9.2$ large-scale eddy turnover times.

\begin{figure}[tb]
  \begin{center}
    \includegraphics[width=\textwidth]{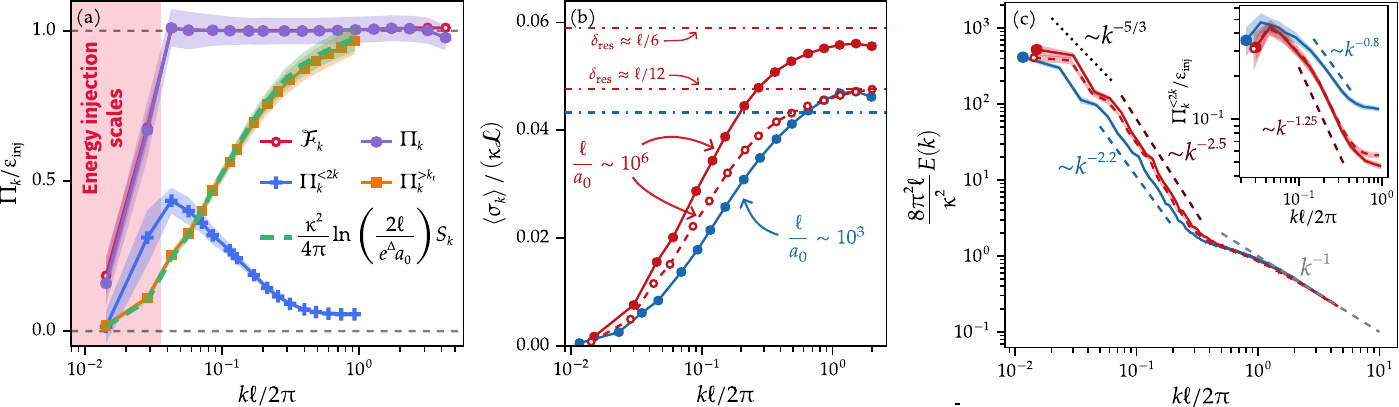}
  \end{center}
  \caption{%
    Influence of line resolution $\deltaRes$ on numerical results.
    (a) Normalized spectral energy fluxes from extra high-resolution run ($\epsInj = 800$, $\azero = 10^{-7}$, $\deltaRes \approx \ell/12$).
    Legend as in Fig.~2 of the main text.
    (b) Mean vortex line production rate $\sigma_k$ due to the coarse-grained velocity $\vs^{<k}$.
    The high-resolution run (dashed red line) is compared to runs A3 ($\epsInj = 800$, $\azero = 10^{-7}$; solid red line) and
    B3 ($\epsInj = 800$, $\azero = 10^{-4}$; solid blue line) of main text, both at $\deltaRes \approx \ell/6$.
    Horizontal dash-dotted lines correspond to $\Csigma = \epsInj / (\kappa \Lcal^2 \tau_\ell)$
    where $\tau_\ell \approx (\kappa / 4\pi) \ln(\ell / a_0)$ is the vortex tension.
    See Fig.~3(b) of main text for more details.
    (c) Kinetic energy spectrum (main panel) and local energy flux (inset).
    Legend as in~(b).
    See Fig.~4 of main text for more details.
  }\label{fig:resolution_effects}
\end{figure}

In Fig.~\ref{fig:resolution_effects} we reproduce the results discussed in the
main text, based this time on the analysis of the high-resolution run.
The scale-by-scale energy fluxes plotted in Fig.~\ref{fig:resolution_effects}(a) display the same qualitative features as those
reported in Fig.~2 of the main text, including a decreasing local energy flux across the inertial scales (blue crosses),
which is compensated by an increasing nonlocal energy transfer towards the quantum scales (orange squares).
Interestingly, the semi-analytical prediction based on $\Scal_k$ (green dashed
line) matches the actual nonlocal transfers even more accurately than in the main text.

In Fig.~\ref{fig:resolution_effects}(b) we compare the normalized mean vortex line production rate $\Lmean{\sigma_k}$
from the high-resolution run (open circles, dashed lines) to those from the
$\epsInj = 800$ runs reported in the main text (filled circles, solid lines).
As hinted in the main text, increasing the line resolution reduces the average
alignment between large-scale velocity gradients and individual vortex lines.
This is likely because the finite line resolution $\deltaRes$ limits the
emergence of high-curvature regions, where the highly fluctuating local vortex
orientations can be expected to decorrelate from the (smooth) large-scale
velocity field.
Note that the resulting value of $\Csigma = \epsInj / (\kappa \Lcal^2
\tau_\ell)$ (dash-dotted horizontal lines), linked in the main text to the
effective viscosity $\nu'$, also decreases at higher resolution, since the vortex
line density $\Lcal$ in the steady state increases for the same $\epsInj$.

Finally, Fig.~\ref{fig:resolution_effects}(c) reports the energy spectrum
obtained from the high-resolution run (dashed lines), again compared with the $\epsInj = 800$
runs of the main text (solid lines).
The energy spectrum only weakly depends on the resolution, especially compared
to the effect of the scale separation $\ell/\azero$ which stays dominant.
This is confirmed by the local energy fluxes plotted in the inset, where the
two curves for $\azero = 10^{-7}$ are nearly indistinguishable.

Overall, these results highlight the robustness of the identified nonlocal energy transfer mechanism
with regards to the numerical line resolution.
Indeed, at high resolution nonlocal transfers persist and are still linked to
production of vortex line by large-scale velocity gradients.
The effect of resolution is mainly visible in the numerical values of
$\Lmean{\sigma_k}$ and $\Csigma$.
Since these values can be related to the effective viscosity $\nu'$, it would
be highly interesting to determine whether they converge as the resolution is
further increased, in which case they may be compared to experimental
estimations near absolute zero~\cite{Walmsley2014,Zmeev2015}.

\subsection{Other numerical parameters}\label{sub:other_numerical_effects}

We also evaluate the influence of other numerical parameters, including the
interpolation method used to evaluate vortex positions and derivatives along
filaments, the simulation time step $\Delta t$, and the critical reconnection
distance relative to the resolution $\deltaRes$.
For this, we perform 4 extra runs with $\epsInj = 200$ and $\azero = 10^{-7}$.
The base run uses the same parameters as the simulations in the main text.
Namely, quintic spline interpolations are used to represent the filament geometry,
the line resolution is $\deltaRes \approx \ell/6$, reconnections are performed
below a critical distance $\deltaRec = \deltaRes$, and the (adaptive) time step is
$\Delta t \approx 0.4 \Delta t_{\text{KW}}$
where $\Delta t_{\text{KW}} = 2 (\deltaRes^2 / \kappa) \left[ \ln(\deltaRes / \pi a_0) + (1/2) - (\Delta + \gamma) \right]^{-1}$
is the period of Kelvin waves with wavelength $\deltaRes$ (see End Matter).
All runs are performed using a 4th-order Runge--Kutta scheme.
The three other runs are performed using:
\begin{enumerate}
  \item cubic spline interpolations (instead of quintic splines);
  \item a smaller time step, $\Delta t \approx 0.2 \Delta t_{\text{KW}}$ (instead of $0.4 \Delta t_{\text{KW}}$), labelled ``$\Delta t / 2$'' in the figures;
  \item a higher resolution, $\deltaRes \approx \ell/12$, but keeping the same
    reconnection distance as in the base case (so that $\deltaRec = 2
    \deltaRes$), labelled ``$\deltaRes / 2$''.
\end{enumerate}
The four runs are started from the same initial condition consisting of
1000 randomly oriented vortex rings.

\begin{figure}[tb]
  \begin{center}
    \includegraphics{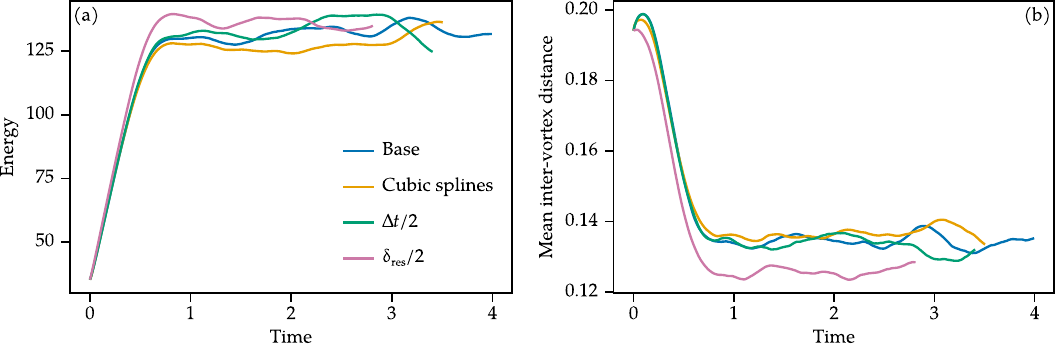}
  \end{center}
  \caption{%
    Influence of numerical parameters on the time evolution of global quantities.
    (a) Kinetic energy.
    (b) Mean inter-vortex distance.
    In all cases, $\epsInj = 200$ and $\azero = 10^{-7}$.
    All runs are started from the same initial condition.
    See text for more details.
  }\label{fig:numerical_time_evolution}
\end{figure}

In Fig.~\ref{fig:numerical_time_evolution} we report the time evolution of
kinetic energy and mean inter-vortex distance $\ell$.
At early times, the total kinetic energy
[Fig.~\ref{fig:numerical_time_evolution}(a)] increases due to the action of the
forcing velocity $\vf$.
This mostly coincides with a decrease of the mean inter-vortex distance
[Fig.~\ref{fig:numerical_time_evolution}(b)] as vortex line is produced.
At later times, both quantities reach statistically-steady states where they
fluctuate around a mean value.
Of all the evaluated parameters, the resolution $\deltaRes$ is the only one to
have a significant impact on both the transient and on the steady-state values,
consistently with Sec.~\ref{sub:resolution_effects}.
Indeed, as $\deltaRes$ is decreased, the numerical dissipation effects detailed
in Sec.~\ref{sec:time_evolution_forcing} also decrease, thus leading to more
energy and more vortex line density for the same energy injection rate.
Interestingly, this is even the case when keeping the same reconnection
distance $\deltaRec$ as in the base run, showcasing as in Sec.~\ref{sec:time_evolution_forcing}
that reconnections only account for a minor part of the total vortex line
losses (and thus numerical energy dissipation) in our simulations.

\begin{figure}[tb]
  \begin{center}
    \includegraphics{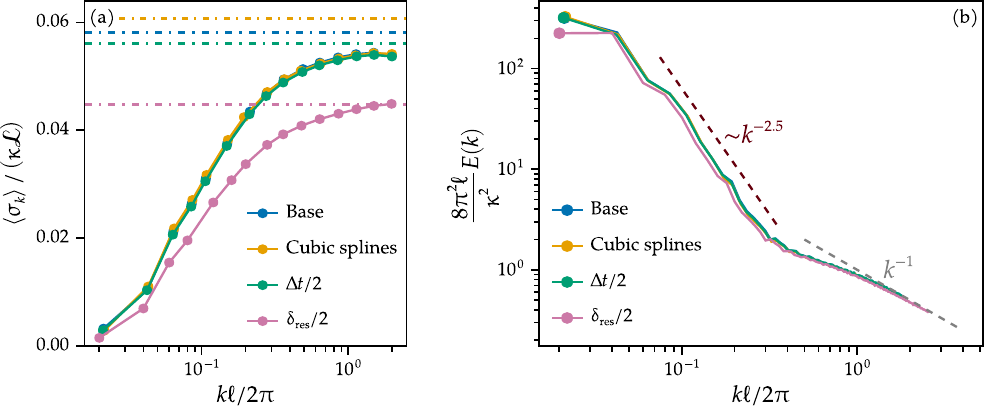}
  \end{center}
  \caption{%
    Influence of numerical parameters on (a) mean vortex line production by coarse-grained velocity gradients and (b) energy spectra.
    In all cases, $\epsInj = 200$ and $\azero = 10^{-7}$.
    In (a), horizontal dash-dotted lines correspond to $\Csigma = \epsInj / (\kappa \Lcal^2 \tau_\ell)$.
    See text for more details.
  }\label{fig:numerical_alignment_spectrum}
\end{figure}

Finally, the influence of these numerical parameters on the results reported in
the main text is evaluated in Fig.~\ref{fig:numerical_alignment_spectrum}.
Reducing the spline order or the time step has no visible impact on either the
normalized mean alignment $\Lmean{\sigma_k}$ between quantum vortices and
large-scale velocity gradients [Fig.~\ref{fig:numerical_alignment_spectrum}(a)]
nor on the resulting energy spectra [Fig.~\ref{fig:numerical_alignment_spectrum}(b)].
In panel (a), a minor influence of these parameters is seen in the values of
$\Csigma$ (dash-dotted horizontal lines), which can be likely associated to the
different average values of $\ell$ in the steady state [see also
Fig.~\ref{fig:numerical_time_evolution}(b)].
These differences may be suppressed with longer simulation times.
On the other hand, and as already discussed in the previous section, the
resolution $\deltaRes$ has a noticeable influence on the normalized values of $\Lmean{\sigma_k}$,
on $\Csigma$, and to a minor extent on the energy spectra.

\section{Shell-to-shell energy transfer matrix}

In the main text we have established the nonlocality of energy transfers from
the positive values of the fluxes $\Pi_{k}^{>k_\ell}$ from large ($< k$) to quantum ($> k_\ell = 2\pi/\ell$) scales, for $k < k_\ell$.
As a more detailed proof of this nonlocality, we consider here the
shell-to-shell energy transfers $T(k, q)$ from scales $k \pm \Delta(k)$ to
scales $q \pm \Delta(q)$.
Here $\Delta(k)$ is the thickness of the shell around wavenumber $k$, roughly corresponding to the distance between two probed wavenumbers $k$.
In the context of the VFM, and in analogy with the definition of $\Pi_k$ in the
main text, shell-to-shell energy transfers can be expressed as
$T(k, q) = \frac{\kappa}{V} \Cint (\svec' \times \vs^{q}) \cdot \vs^{k} \, \dd\xi$,
where $\vs^{k}$ and $\vs^{q}$ are the superfluid velocities band-pass filtered around wavenumbers $k$ and $q$, respectively.
It is trivial to show that this quantity is antisymmetric, i.e.\ $T(q, k) = -T(k, q)$, and in particular $T(k, k) = 0$.
In the case of a predominantly local cascade, as in classical 3D turbulence, and for $k$ in the inertial range,
one would expect $T(k, q)$ to be positive for $q \gtrsim k$ (slightly below the
diagonal) and zero for $q \gg k$.

\begin{figure}
  \begin{center}
    \includegraphics[width=\textwidth]{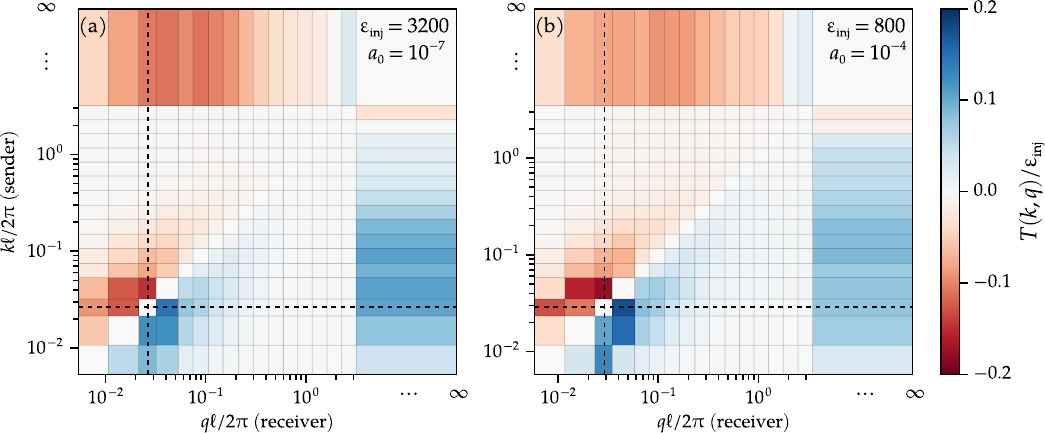}
  \end{center}
  \caption{%
    Shell-to-shell energy transfer matrix $T(k, q)$ from two different forced runs (see Table I in main text).
    Values are normalized by the energy injection rate $\epsInj$ in each run.
    Results have been averaged across (a) 40 and (b) 15 flow snapshots within the statistically steady state.
    Black dashed lines correspond to the forcing scale ($\kf = 2.5$).
    In each plot, the rightmost column and the uppermost row include wavenumbers from $k \approx 3 \kell$ up to $k = \infty$.
  }\label{fig:energy_transfer_matrix}
\end{figure}

In Fig.~\ref{fig:energy_transfer_matrix} we report shell-to-shell energy
transfer matrices from two forced runs, normalized by the mean energy injection
rate $\epsInj$ in each simulation.
In the following we focus on the lower-right half of the plots (the upper-left half is redundant
due to the antisymmetry of $T$).
At scales slightly smaller than the forcing one (dashed black lines), values of
$T$ are slightly positive below the diagonal, confirming that part of the
energy cascades locally to smaller scales.
However, these values clearly decrease as larger wavenumbers are considered.
In other words, the local energy flux decreases with $k$, consistently with
the monotonously decreasing behavior of $\Pi_k^{<2k}$ for $k > \kf$ (Fig.~2 in main text).
Note that, in the case $a_0 = 10^{-4}$, near-diagonal values are slightly larger than
for $a_0 = 10^{-7}$, confirming that a larger $\ell/a_0$ tends to further
suppress the local cascade (as also seen in the inset of Fig.~4 in the main text).

More strikingly, the rightmost columns of the two transfer matrices $T(k, q)$ in
Fig.~\ref{fig:energy_transfer_matrix} display predominantly positive values.
These columns include wavenumbers from $q \approx 3 k_\ell$ (chosen
arbitrarily) to $q = \infty$.
Hence, in this case the ``receiving`` velocity $\vs^q$ is the high-pass
filtered velocity at quantum scales $q \gtrsim k_\ell$.
As a result, the positive values of the rightmost columns are the signature of
a nonlocal energy transfer from inertial scales $k \ll k_\ell$ to the quantum scales,
completely bypassing the intermediate scales.
Importantly, the values of the nonlocal fluxes are of the same order (and often
larger) than the near-diagonal ones associated to a local energy flux.
This is fully consistent with the behavior of the nonlocal transfer $\Pi_k^{>k_\ell}$
(see Fig.~2 of the main text).
Overall, these results confirm the picture of a predominantly nonlocal energy
transfer towards quantum scales discussed in the main text.

\section{Compensated energy spectra}

\begin{figure}
  \begin{center}
    \includegraphics{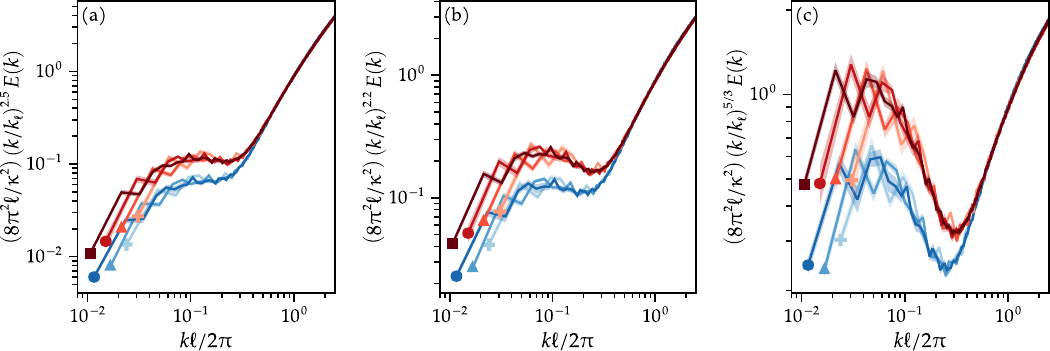}
  \end{center}
  \caption{%
    Compensated kinetic energy spectrum from different runs.
    Energy spectra are compensated by (a) $(k / k_\ell)^{2.5}$, (b) $(k / k_\ell)^{2.2}$ and (c) $(k / k_\ell)^{5/3}$.
    Two sets of curves are shown with $\ell / \azero \sim 10^{6}$ (red) and $10^{3}$ (blue).
    Colors and markers are the same as in Fig. 4 of the main text.
    Light-colored bands indicate 95\% confidence intervals.
  }\label{fig:compensated_spectra}
\end{figure}

As a complement of the energy spectra in Fig. 4 of the main text, in
Fig.~\ref{fig:compensated_spectra} we plot the same energy spectra compensated by different power laws.
Panel (a) confirms that the $\ell / a_0 \sim 10^{6}$ simulations (red curves)
display a small wavenumber region $0.06 \lesssim k / k_\ell \lesssim 0.2$,
where the spectrum can be roughly fitted by a power law scaling $E(k) \sim k^{-2.5}$.
Similarly, panel (b) confirms that the energy spectrum in the
$\ell / a_0 \sim 10^{3}$ cases (blue curves) roughly behaves as $E(k) \sim k^{-2.2}$ within the same (limited) wavenumber region.
Finally, Fig.~\ref{fig:compensated_spectra}(c) shows the energy spectra compensated by the Kolmogorov-like scaling $E(k) \sim k^{-5/3}$,
further supporting that our spectra significantly deviate from this scaling law for $k/k_\ell \gtrsim 0.05$.
Interestingly, at smaller wavenumbers, the spectra suggest the appearance of a
plateau hinting at the recovery of quasi-classical scaling laws at large scales.
This would be consistent with the expectation that
nonlocal energy fluxes -- and the alignment between velocity and quantum
vortices -- should vanish as larger scales are considered, as discussed in the main text.
However, an even larger scale separation between the forcing and the quantum
scales -- currently out of reach for VFM numerical simulations -- is needed to
confirm this hypothesis.
We insist that the main conclusion from these results is that they
showcase significant spectral steepening compared to the expected
Kolmogorov-like spectrum, over a significant (but possibly limited) range of
scales larger than $\ell$, and that this effect is amplified by the scale
separation between the quantum scale $\ell$ and the vortex core size $a_0$.

\section{Decaying quantum turbulence simulations}

\begin{figure}
  \begin{center}
    \includegraphics[width=\textwidth]{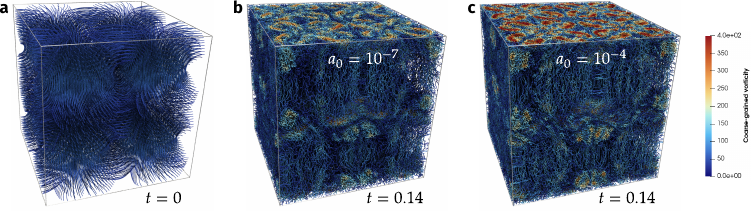}
  \end{center}
  \caption{Visualization of decaying Taylor--Green quantum turbulence simulations.
    (a)~Quantum vortices at $t = 0$ and (b-c) at $t = 0.14$, for vortex core radius (b) $a_0 = 10^{-7}$
    and (c) $a_0 = 10^{-4}$.
    Colors represent the Gaussian-filtered vorticity at the scale of the instantaneous mean inter-vortex distance $\ell$.
  }\label{fig:TG_visualisation}
\end{figure}

To confirm the robustness of the identified nonlocal energy transfer mechanism,
in the following we report decaying VFM numerical simulations
starting from the standard Taylor--Green initial condition~\cite{Nore1997,Araki2002},
in the absence of the energy injection term $\vf$.
Quantum vortices are initialized so that they are
oriented parallel to the vorticity $\vortvec = \curl \vvec$ associated to
the Taylor--Green velocity field, $\vvec(x, y, z) = \left[ \sin(x) \cos(y) \sin(z), -\cos(x) \sin(y) \sin(z), 0 \right]$.
The initial mean inter-vortex distance is $\ell_0 \approx 0.148$ and the  
line resolution is set to $\delta_{\text{res}} \approx 0.0118$  
(the periodic domain size is $\Lbox = 2\pi$),
ensuring that $\ell / \delta_{\text{res}} > 5$ through all simulations.
To evaluate the influence of the $\ell / a_0$ ratio, we perform two different
runs with the vortex core radius set to $a_0 = 10^{-7}$ and $a_0 = 10^{-4}$ as in the main text.
The initial condition, visible in Fig.~\ref{fig:TG_visualisation}(a), consists of
2112 vortex filaments discretized with $N \approx 7.3 \times 10^5$ vortex points.

\subsection{Time evolution of global quantities}

\begin{figure}
  \centering
  \includegraphics{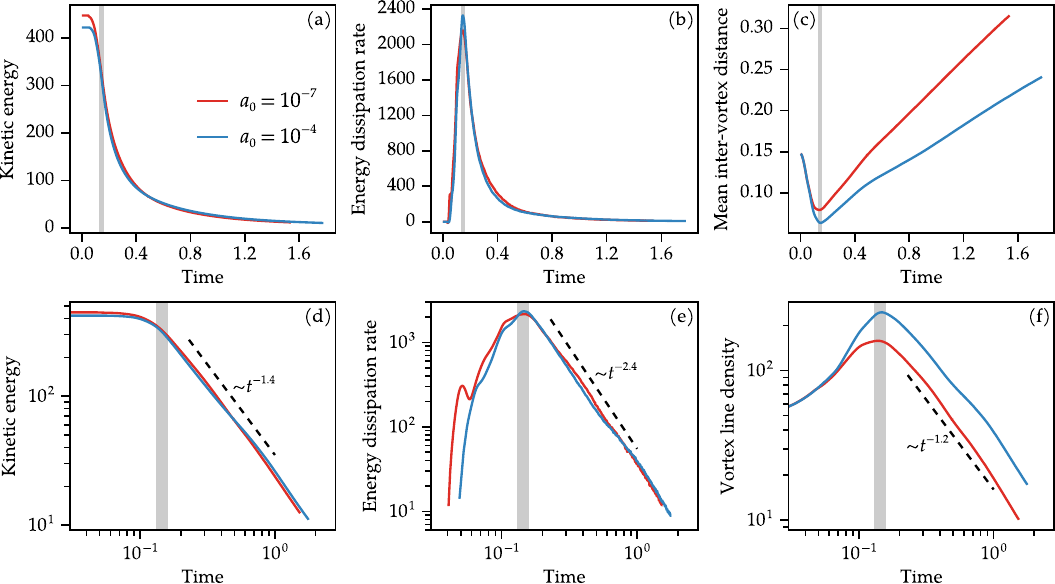}
  \caption{%
    Temporal decay of Taylor--Green quantum turbulence simulations.
    (a) Total kinetic energy $E(t)$, (b) energy dissipation rate $\varepsilon(t) = -\frac{\dd E(t)}{\dd t}$, and (c) mean inter-vortex distance $\ell(t)$
    in linear scale;
    (d) Total kinetic energy, (e) energy dissipation rate and (f) vortex line density $\Lcal(t)$ in logarithmic scale.
    Quantities are plotted for $a_0 = 10^{-7}$ (red curves) and $10^{-4}$ (blue curves).
    In each panel, the gray region represents the time span over which statistics are gathered,
    at which energy dissipation and vortex line density are close to their
    maximal values.
  }\label{fig:TG_decay}
\end{figure}

Figure~\ref{fig:TG_decay} reports the time evolution of relevant global quantities.
Initially, despite the vortex configurations being identical, the case $a_0 =
10^{-7}$ has a higher energy content [Fig.~\ref{fig:TG_decay}(a)] than the case $a_0 = 10^{-4}$.
This is simply explained by the wider span of the quantum scales $\ell / a_0$, to which is associated a small-scale energy
$E^{>k_\ell} \sim (\kappa^2 / 4\pi\ell^2) \ln\left( \ell / a_0 \right)$.
At later times, following a short transient, energy decays monotonically over
time as reconnections take place.
The energy dissipation rate $\varepsilon$ [Fig.~\ref{fig:TG_decay}(b)], directly estimated as the time derivative of kinetic energy,
reaches a maximum near $t \approx 0.14$, which also roughly corresponds to the
time where the mean inter-vortex distance $\ell$ [Fig.~\ref{fig:TG_decay}(c)] reaches its minimal value.
Before this, $\ell$ decreases over time (i.e.\ the vortex line density $\Lcal = \ell^{-2}$
increases), which, according to our present findings, can be interpreted as the
production of vortex line by the large-scale self-induced superfluid velocity.
As pointed out in the main text, the non-monotonous behavior of $\varepsilon$ and $\ell$ is a common feature of
zero-temperature quantum turbulence simulations using either the
VFM~\cite{Araki2002} or the Gross--Pitaevskii model~\cite{Nore1997,Villois2016a,Muller2020} -- which is more relevant at
near-atomic scales.
In particular, the time where dissipation and vortex line density reach their
maximal values is generally interpreted as a quasi-steady state where the flow
is traditionally expected display Kolmogorov-like features (see e.g.~\cite{Muller2020}).

The influence of $\ell / a_0$ is manifest in Fig.~\ref{fig:TG_decay}.
In particular, $\ell$ reaches a smaller minimal value for the $a_0 = 10^{-4}$ case
($\ell / a_0 \sim 10^{3}$, blue curves) than for the $a_0 = 10^{-7}$ case ($\ell / a_0 \sim 10^{6}$, red curves).
This is likely explained by the slower depletion of the large-scale energy by nonlocal transfers associated to a smaller value
of $\ell / a_0$ [see Eq.~(5) in main text].
As a result, the large-scale flow has more time to produce vortex length resulting in a smaller minimal value of $\ell$.
Consistently, since the vortex line density is larger, the $a_0 = 10^{-4}$ case also reaches a higher energy dissipation rate (which is
associated to vortex line losses, see Sec.~\ref{sec:time_evolution_forcing}).
The difference in vortex line density between these two cases is also visually apparent
in Fig.~\ref{fig:TG_visualisation}(b-c), where
the instantaneous vortex states near the maximum dissipation time are shown.
Note that these maximum dissipation states retain much of the large-scale anisotropy of the Taylor--Green initial condition.
This contrasts with the isotropy of the forced simulations in the main text --
where the initial condition consists of randomly oriented vortex rings and the forcing has no preferential orientation.

In quasi-classical quantum turbulence, the vortex line density is expected to decay as $\Lcal \sim t^{-3/2}$.
This scaling arises from the expectation that energy is dissipated at a rate $\varepsilon(t) = -\frac{\dd E}{\dd t} = \nu' (\kappa \Lcal)^2$
and that $\varepsilon(t) \sim t^{-3}$ in decaying classical and quasi-classical turbulence~\cite{Vinen2002},
which also results in $E(t) \sim t^{-2}$.
In Fig.~\ref{fig:TG_decay}(d-f) we show that these scalings are not found in
our decaying simulations, contrarily to existent experimental evidence in \HeII{}
near absolute zero~\cite{Walmsley2007,Zmeev2015}.
This is likely explained by the finite span of the inertial range in our
simulations and by the predominance of nonlocal energy transfers, as confirmed
in the following section.

\subsection{Energy spectra and scale-by-scale energy fluxes}

\begin{figure}
  \centering
  \includegraphics{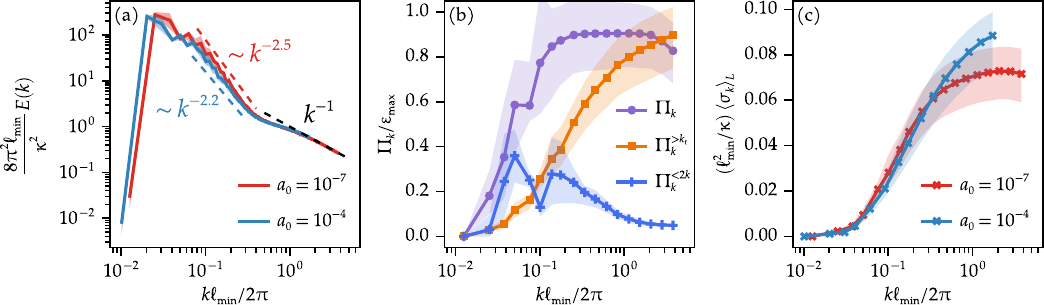}
  \caption{%
    Energy spectra and spectral energy fluxes in decaying Taylor--Green quantum turbulence simulations.
    (a) Normalized energy spectra for $a_0 = 10^{-7}$ (red) and $10^{-4}$ (blue).
    (b) Total energy flux $\Pi_k$ (circles), local energy flux $\Pi_k^{<2k}$
    (crosses) and nonlocal energy flux to quantum scales $\Pi_k^{>k_\ell}$
    (squares) obtained in the $a_0 = 10^{-7}$ case.
    (c) Normalized mean vortex line production rate $\Lmean{\sigma_k}$
    associated to the coarse-grained velocity field $\vs^{<k}$ [Eq.~(6) in main text].
    Values are normalized using the minimal mean inter-vortex distance $\ell_{\min}$ associated to each case [Fig.~\ref{fig:TG_decay}(c)].
    Spectra and fluxes are averaged from 30 samples within the time window $t \in [0.13, 0.16]$ (gray area in Fig.~\ref{fig:TG_decay}).
    Light-colored bands correspond to fluctuations within the chosen temporal window.
  }\label{fig:TG_spectra_flux}
\end{figure}

The kinetic energy spectrum near the maximum dissipation state is shown in
Fig.~\ref{fig:TG_spectra_flux}(a) for the two considered values of $a_0$.
Overall, the spectra in the decaying runs display the same main features as
those obtained in the forced cases [see Fig.~(3) in main text].
The $k^{-1}$ regime is clearly visible in the quantum scales ($k \ell \gg 1$).
At larger scales, the two cases display spectra that are considerably steeper
than the expected Kolmogorov-like scalings ($E(k) \sim k^{-5/3}$).
As in the forced simulations, this behavior is more dramatic for the larger value of $\ell / a_0$,
which is explained by more intense nonlocal energy transfers from large to quantum scales.
In both cases, one can identify power law scalings over a small wavenumber region, with exponents that are
quantitatively consistent with those found in the forced cases ($-2.5$ and $-2.2$).

The spectral energy fluxes are considered in more detail in Fig.~\ref{fig:TG_spectra_flux}(b) for the case $a_0 = 10^{-7}$.
The total flux $\Pi_k$ across wavenumber $k$ (circles) is nearly constant over about one decade, in a region that includes the
quantum scale $k_\ell = 2\pi / \ell$.
This behavior is qualitatively similar to that observed in the forced
simulations [Fig. 2 in the main text], although in those simulations, steady
energy injection at large scales enables a constant energy flux over a much
wider range of scales.
As in the forced cases, the nonlocal flux $\Pi_k^{>k_\ell}$ to quantum scales
monotonously increases with $k$, indicating that the energy cascade is depleted
via nonlocal transfers through all scales.
Conversely, the local flux $\Pi_k^{<2k}$ does not display a plateau as would be
expected for a Kolmogorov-like energy cascade.
Instead, it decreases with $k$ in the constant $\Pi_k$ region as the cascade is depleted.
Finally, Fig.~\ref{fig:TG_spectra_flux}(c) displays the average vortex line
production rate $\Lmean{\sigma_k}$ associated to the low-pass filtered velocity $\vs^{<k}$.
The obtained values are positive at all scales and increase as smaller scales are taken into account,
indicating that, on average, all scales produce vortex line density (and thus transfer energy to quantum scales).
When normalized by $\kappa$ and $\ell$, the stretching rate displays values
which are quantitatively consistent with those of our forced simulations [Fig.~(3b) in main text],
despite the temporal variations and statistical convergence issues inherent to the decaying configuration.
Overall, these results are fully consistent with those obtained from forced simulations in the main text.
Moreover, they neatly illustrate the usefulness of a forcing scheme confined over a controlled range of scales
for obtaining a constant energy flux across a wide range of scales.

\end{document}